\documentclass[11pt]{article}

\usepackage{arxiv}
\usepackage{graphicx}
\usepackage{natbib}
\usepackage{url} 
\usepackage{amssymb}
\usepackage{amsmath}
\usepackage{lmodern}
\usepackage{mathrsfs}%
\usepackage{subcaption}
\usepackage[title]{appendix}%
\usepackage{xcolor}%
\usepackage{booktabs}%
\usepackage[ruled]{algorithm2e}
\usepackage{listings}
\usepackage{amsfonts}
\usepackage{amsthm}
\usepackage{mathtools}
\mathtoolsset{showonlyrefs=true}
\usepackage{subcaption}
\usepackage{multirow}
\usepackage{bm}
\usepackage{float}
\usepackage{adjustbox}

\newcommand{\bolds}[1]{\boldsymbol{#1}}

\newcommand{\calE}{{\cal E}}

\newcommand{\calG}{{\cal G}}

\newcommand{\calL}{{\cal L}}

\newcommand{\calP}{{\cal P}}

\newcommand{\calV}{{\cal V}}
\newcommand{\calW}{{\cal W}}

\newcommand{\calZ}{{\cal Z}}

\newcommand{\bB}{\mathbf{B}}

\newcommand{\bbE}{\mathbb{E}}

\newcommand{\bI}{\bolds{I}}

\newcommand{\bq}{\bolds{q}}
\newcommand{\bQ}{\mathbf{Q}}

\newcommand{\bbR}{\mathbb{R}}

\newcommand{\bS}{\mathbf{S}}

\newcommand{\bV}{\mathbf{V}}

\newcommand{\bW}{\bolds{W}}
\newcommand{\bx}{\bolds{x}}

\newcommand{\by}{\bolds{y}}
\newcommand{\bY}{\bolds{Y}}
\newcommand{\bz}{\bolds{z}}
\newcommand{\bZ}{\mathbf{Z}}

\newcommand{\IdentityMat}{\bI}
\newcommand{\Norm}{\mathcal{N}}
\newcommand{\bzero}{\mathbf{0}}

\newcommand{\boeta}{\bolds{\eta}}

\newcommand{\bSigma}{\bolds{\Sigma}}

\newcommand{\bmu}{\bolds{\mu}}

\newcommand{\bLambda}{\bolds{\Lambda}}

\newcommand{\bpsi}{\bolds{\psi}}
\newcommand{\bPsi}{\bolds{\Psi}}
\newcommand{\bphi}{\bolds{\phi}}

\newcommand{\given}{\,|\,}
\newcommand{\lrnd}{\left(}
\newcommand{\rrnd}{\right)}
\newcommand{\lsq}{\left[}
\newcommand{\rsq}{\right]}
\newcommand{\lcur}{\left\lbrace}
\newcommand{\rcur}{\right\rbrace}
\renewcommand{\tilde}{\widetilde}

\newcommand{\tCollapsedurl}{\if0\blind
	{
		\url{\tCollapsedurl}
	} \fi
	\if1\blind
	{ 
		\url{github.com/}\texttt{[url redacted in blinded version]}
	} \fi}

\title{A new hierarchical distribution on arbitrary sparse precision matrices}

\author{
Gianluca Mastrantonio\\
    \scriptsize{Dipartimento di Scienze Matematiche}\\
    \scriptsize{Politecnico di Torino}\\
    \scriptsize{\texttt{gianluca.mastrantonio@polito.it}}\\
    \And
Pierfrancesco Alaimo Di Loro\\
    \scriptsize{Dipartimento GEPLI}\\
    \scriptsize{Libera Università Maria Ss. Assunta (LUMSA)}\\
    \scriptsize{\texttt{p.alaimodiloro@lumsa.it}}\\
\And
    Marco Mingione\\
    \scriptsize{Dipartimento di Scienze Politiche}\\
    \scriptsize{Università Roma Tre}\\
    \scriptsize{\texttt{marco.mingione@uniroma3.it}}\\
}

\begin{document}
\maketitle
\begin{abstract}
We introduce a general strategy for defining distributions over the space of sparse symmetric positive definite matrices. Our method utilizes the Cholesky factorization of the precision matrix, imposing sparsity through constraints on its elements while preserving their independence and avoiding the numerical evaluation of normalization constants. In particular, we develop the \textit{S-Bartlett} as a modified Bartlett decomposition, recovering the standard Wishart as a particular case.  By incorporating a Spike-and-Slab prior to model graph sparsity, our approach facilitates  Bayesian estimation through a tailored MCMC routine based on a Dual Averaging Hamiltonian Monte Carlo update. This framework extends naturally to the Generalized Linear Model setting, enabling applications to non-Gaussian outcomes via latent Gaussian variables.
We test and compare the proposed S-Bartelett prior with the G-Wishart both on simulated and real data.   
Results highlight that the S-Bartlett prior offers a flexible alternative for estimating sparse precision matrices, with potential applications across diverse fields.
\end{abstract}

\section{Introduction}
Accounting for dependence structure across observations is pivotal when dealing with spatio-temporal applications or multivariate outcomes. For Gaussian random variables, this dependence is fully characterized by the covariance matrix and its inverse, known as the precision matrix. While covariance matrix elements capture the marginal dependency between observations, those of the precision matrix describe their conditional dependence. The conditional dependencies structure can be elegantly represented using graph structures, often referred to as \textit{graphical models} \citep{Lauritzen1996}. 

When the conditional dependence structure across observations is known a priori, as is common with spatial or temporal data, the sparsity of the precision matrix can be exploited to define parsimonious models and develop efficient estimation algorithms \citep{rue2009, datta2016, Katzfuss202017}.
However, when no prior information about dependence is available, the Wishart distribution naturally arises as a distribution over the space of precision matrices for Gaussian random variables, playing a significant role as prior in the context of Bayesian statistics. Being defined on a continuum, it is incapable of producing estimates with exact zeroes. However, the data might still exhibit a conditional independence structure even when it is unknown a-priori. Most importantly, this unknown pattern is of major interest in many application fields like ecology \citep{peruzzi2024spatial}, genomics \citep{opgen2006inferring}, economics \citep{cserban2007modelling}.
Various methods for estimating sparse precision matrices have been developed under both frequentist and Bayesian paradigms. Frequentist approaches, such as node-wise regression \citep{Zhou} and graphical lasso \citep{friedman2008sparse, Avagyan}, are computationally efficient but lack probabilistically consistent measures of uncertainty regarding the graph structure.  Conversely, Bayesian methods overcome the above-mentioned issue, offering valuable probabilistic insights. They are mostly based on the G-Wishart distribution \citep{roverato2002hyper}, which yields interpretable and coherent uncertainty measures.
The G-Wishart can be viewed as a hierarchical prior: a distribution is assumed over a latent graph $\calG$ representing the unknown conditional dependence structure and, conditionally on this graph, the precision matrix density is taken as proportional to that of a Wishart distribution. A major practical challenge lies in evaluating the normalization constant of this conditional density, which involves integrating over all Wishart matrix elements consistent with the conditioning graph structure. Since this integral generally lacks a closed-form solution, exact evaluation must be replaced by numerical approximations.
In response, 
extensive research has focused on improving the computational efficiency of the G-Wishart prior, including strategies for approximating or evaluating its normalization constant \citep{lenkoski2011computational, Wang2015, Sagar2021,vandenBoom2021}.
Among these advancements, the implementation proposed by \cite{Willem} leverages the Gaussian distribution of the outcomes to effectively harness the full potential of the G-Wishart distribution within an efficient estimation framework.

In this paper, we propose an alternative strategy to define distributions over the space of sparse symmetric positive definite matrices. We specify a hierarchical prior named \textit{S-Bartlett} (Sparse-Bartlett), which circumvents the need for numerically evaluating normalization constants.
Rather than working directly with elements of a Wishart-distributed precision matrix, the S-Bartlett prior leverages the Cholesky factorization of a matrix and mimics the Bartlett decomposition to find suitable distributions over its elements. 
By enforcing specific constraints, exact zeroes are induced in the corresponding position of the precision matrix, ensuring sparsity. At the same time, the parameters of the original Bartlett random variable are adjusted to account for the above-mentioned sparsity constraints.
We incorporate this conditional distribution with a Spike and Slab prior \citep{mitchell1988bayesian} to select the edges in the graphs (i.e., the $0$s in the precision matrix). 
For implementation, we develop an MCMC routine combining Hamiltonian Monte Carlo (HMC, specifically the No-U-Turn Sampler) for updating the Cholesky factor and Gibbs sampling for the Spike-and-Slab indicator variables. 
This general estimation method seamlessly extends to the \textit{Generalized Linear Model} framework, enabling applications to non-Gaussian outcomes through latent Gaussian variables.

\section{Graphical models and precision matrices} \label{sec:graphical_model}
Let $\by_i,\, i=1,\dots n$ be independent realizations of multivariate random variables $\bY_i\in\bbR^p$. The conditional dependence structure among the elements of $\bY_i$ is typically represented in graphical modeling through an undirected graph $\calG=\lrnd \calV, \calE\rrnd$, where $\calV=\lcur 1,\dots, p\rcur$ is the set of nodes corresponding to each element of $\bY_i$, and $\calE\subseteq\lcur (j, k)\,:\, 1\leq j < k\leq p\rcur$ is a set of edges representing links among them. Let $N(j)=\lcur k\,:\, (j,\,k)\in\calV\rcur\subset\calV$ define the set of neighbors induced by $\calG$ over element $j$. The links define the conditional dependence relationships between each pair of nodes, specifically $Y_{ij}=\lsq\bY_i\rsq_j$ is conditionally independent of all elements in $\calV\setminus N_j$ given its neighbors $\bY_{i, N(j)}$. This property can be expressed through density functions as:
\begin{equation*}
    f_{Y_{ij}\given \bY_{i,-j}}\lrnd y_{ij}\given \by_{i,-j} \rrnd=f_{Y_{ij}\given \bY_{i,N(j)}}\lrnd y_{ij}\given \by_{i,N(j)} \rrnd,
\end{equation*}
where $\bY_{i, -j}=\lsq Y_{i1},\dots,Y_{i,j-1}, Y_{i,j+1}, \dots, Y_{ip}\rsq$.
In Gaussian Graphical Models (GGM), the random vectors are assumed to follow a multivariate Gaussian distribution $\bY_i\, \overset{ind}{\sim}\, \Norm\lrnd\bmu_i,\, \bSigma\rrnd$, where $\bmu_i$ is the $p$-dimensional mean vector of the $i$-th observation and $\bSigma$ is a common $p \times p$ symmetric positive definite covariance matrix.
Collecting the $n$ vectors in the $n \times p$ matrix $\mathbf{Y} = \lsq \bY_1, \bY_2, \dots , \bY_n \rsq^{\top}$, their joint density is:
\begin{equation}
\label{eq:jdensNorm}
    f_{\bY}(\bY\given \bmu,\, \bSigma) = \lrnd 2 \pi\rrnd^{-\frac{n\cdot p}{2}}\cdot {\left| \bSigma^{-1} \right|}^{\frac{1}{2}} \cdot \exp \lcur -\frac{1}{2}\cdot\sum_{i=1}^n (\by_i-\bmu_i)^{\top}\bSigma^{-1}(\by_i-\bmu_i)\rcur,
\end{equation}
where $\boldsymbol{\mu} = \lsq\boldsymbol{\mu}_1, \boldsymbol{\mu}_2, \dots , \boldsymbol{\mu}_n\rsq^{\top}$. Notice that the density in Equation \eqref{eq:jdensNorm} depends on $\bSigma$ only through its inverse, $\bLambda = \bSigma^{-1}$, known as the precision matrix. Hence, it can be reparametrized as $f_{\bY}(\bY\given \bmu,\, \bLambda)$. While $\bSigma$ describes the marginal variances and covariances of the elements of $\bY_i$, the precision matrix $\bLambda$ represents their conditional variances. For Gaussian vectors, $\bLambda$ fully encodes the conditional dependence structure, establishing a dual relationship between the edges in $\calG$ and the non-zero elements in $\bLambda$. Consequently, recovering the unknown structure of $\calG$ to understand the conditional dependence relationships across the elements in $\bY_i$ is equivalent to identifying the sparsity pattern in $\bLambda$. However, this requires exploring the space of positive definite matrices adhering to arbitrary sparsity patterns, which poses significant challenges.

\subsection{The Wishart and G-Wishart distribution}
\label{subsec:wishart}

Before defining a distribution over the space of sparse positive definite matrices, we review the $p$-dimensional Wishart distribution. It is the most natural choice for modeling distributions over the space of arbitrary $p\times p$ positive definite matrices, say $\calP_p$, serving as a conjugate prior to the precision matrix of multivariate Gaussian random variables in Bayesian statistics.  A $p\times p$ matrix $\bLambda$ is said to follow a Wishart distribution with $\tilde{\nu}>0$ degrees of freedom and positive definite scale matrix $\bS$ if $\boldsymbol{\Lambda} \sim \calW\lrnd\tilde{\nu}, \mathbf{S}\rrnd$. In particular, this distribution maps only positive definite matrices for $\tilde{\nu}\geq p$ and, considering our interest in this subset only, we can reparametrize the distribution by $\nu=\tilde{\nu}-p>0$ \citep{Willem}. The corresponding probability density function is given by:
\begin{equation*}
f_{\calW}\lrnd\bLambda\given \nu, \bS\rrnd=\frac{\left|\bLambda\right|^{(\nu-1)/2}}{\lrnd 2^{p}\cdot\left|\bS\right|\rrnd^{(\nu+p)/2}\cdot\Gamma_p((\nu+p)/2)}\cdot \exp\lcur-\frac{1}{2}\cdot \text{tr}\lrnd\bS^{-1}\bLambda\rrnd\rcur,
\end{equation*}
where $\Gamma_p(\cdot)$ denotes the multivariate Gamma function. 

The G-Wishart distribution, first introduced in \cite{roverato2002hyper}, characterizes positive definite matrices adhering to the sparse precision pattern induced by a given graph $\calG=\lrnd\calV, \calE\rrnd$. Since its inception, it has gained significant attention as a sparsity-inducing prior in GGMs, as evidenced by a substantial body of research (see e.g. \cite{mohammadi2015bayesian, Willem}, and references therein).
This distribution is defined conditionally on the zeroes induced by the graph $\calG$ and has support on $\calP_{p}(\calG)\subseteq\calP_p$, i.e. the space of $p\times p$ positive-definite matrices satisfying $[\bLambda]_{ij}=0$ for all $(i,j)\notin\calE$. Specifically, we denote $\bLambda\given\calG, \nu \sim \calG\calW\lrnd\nu, \bS\rrnd,\, \bLambda\in\calP_p\lrnd\calG\rrnd$, a matrix $\bLambda$ that follows the G-Wishart distribution.
Its density is defined as being proportional to that of the classical Wishart, but restricted to the space of $\calP_{p}(\calG)$:
\begin{equation}
\label{eq:GWishdens}
    f_{\calG\calW}\lrnd\bLambda\given \calG, \nu,\bS\rrnd=\frac{f_{\calW}\lrnd\bLambda\given \nu, \bS\rrnd}{\int_{\calP_p\lrnd\calG\rrnd}f_{\calW}\lrnd\bLambda\given \nu, \bS\rrnd d\bLambda}.
\end{equation}
Although the definition of the G-Wishart density is straightforward, its practical use poses significant computational challenges. The integral that serves as a normalization constant in Equation \eqref{eq:GWishdens} lacks a closed-form solution. This is not problematic when performing inference on $\bLambda$ for a given (known) graph $\calG$ as \eqref{eq:GWishdens} can be evaluated up to proportionality constant non depending on $\bLambda$. However, its evaluation becomes essential whenever the graph $\calG$ is unknown and inference on the conditional dependence structure is sought. Many strategies have been proposed to approximate such quantity, each striking different balances between accuracy and computational efficiency.

\subsection{The Bartlett decomposition}
\label{subsec:bartlett}
Wishart-distributed random matrices with scale matrix $\bS$ can be expressed as transformations of Wishart-distributed random matrices with identity scale matrix $\IdentityMat_p$. Let $\boldsymbol{\Psi}$ denote the lower triangular matrix obtained via the Cholesky decomposition of $\mathbf{S} = \bPsi\bPsi^{\top}$. 
Then, $\bLambda \sim \calW(\nu, \mathbf{S})$ if and only if $\bLambda=\bPsi\tilde{\bLambda}\bPsi^{\top}$, where $\tilde{\bLambda}\sim\calW\lrnd\nu, \IdentityMat_p\rrnd$.
Further, if we consider the Cholesky factorization of $\tilde{\bLambda}$ into the product of a lower-triangular matrix $\bB$ and it transpose, it follows that $\bLambda = \bPsi\mathbf{B} \mathbf{B}^{\top}\bPsi^{\top}$. This is known as \textit{Bartlett decomposition} and it facilitates the characterization of the distribution of $\bLambda$ in terms of lower-dimensional independent distributions over the elements of $\bB$. In particular,
\begin{equation}
\label{eq:distBart} 
\tilde{\bLambda}\sim \calW\lrnd \nu, \IdentityMat_p\rrnd \, \iff \, 
\begin{aligned}
    b_{kk}^2 &\overset{ind}{\sim} \chi^2(\nu+p - k),\quad k=1,\dots,p,\\
    b_{jk} &\overset{ind}{\sim} \Norm(0,\,1),\quad j=2,\dots,p,\; k=1,\dots, j-1.
\end{aligned} 
\end{equation}
Note that degrees of freedom of the diagonal elements can be interpreted as the sum of a base degree $\nu>0$ and the number of non-zero off-diagonal elements in column $k$.


The Bartlett decomposition can be extended to characterize the elements of $\bQ = \bPsi\bB$, which is the lower triangular matrix corresponding to the Cholesky decomposition of the original $\bLambda$, i.e. $\bLambda = \bQ \bQ^{\top}$. 
The individual elements $q_{jk}=\lsq\bQ\rsq_{jk}$ are given by:
\begin{equation}
\label{eq:q}
  q_{jk} = 
  \begin{cases}
    \psi_{kk}\cdot b_{kk} & \text{if } j = k\\
    \psi_{jk}\cdot b_{kk} + \sum_{t=k+1}^j \psi_{jt}\cdot b_{tk} & \text{if } j > k\\
    0 & \text{otherwise}.
\end{cases}
\end{equation}
Using the expression of Equation \eqref{eq:q} together with Equation \eqref{eq:distBart}, the distribution of the elements of $\bQ$ can be derived. The diagonal elements are independently Gamma-distributed random variables:
\begin{equation}
    \label{eq:qdistdiag}
    q_{kk}^2 \overset{ind}{\sim} Ga\lrnd\frac{\nu+ p - k}{2},\, \frac{1}{2\cdot\psi_{kk}^2} \rrnd, \quad k = 1, \dots, p.
\end{equation}
The non-zero off-diagonal elements (i.e. those below the diagonal), conditioned on the diagonal elements $q_{kk}$, are column-wise independent Gaussian random vectors:
\begin{equation}
        \label{eq:qdistoffdiag}
    \mathbf{q}_{k}\given q_{kk} \sim \Norm_{p-k} \lrnd\frac{q_{kk}}{\psi_{kk}} \cdot \boldsymbol{\psi}_{k},\, \boldsymbol{\Psi}_{kk} \boldsymbol{\Psi}^{\top}_{kk} \rrnd,
\end{equation}
where $\mathbf{q}_{k}=\mathbf{Q}_{k+1:p, k}$, $\boldsymbol{\psi}_{k}=\boldsymbol{\Psi}_{k+1:p, k}$, and $\boldsymbol{\Psi}_{kk}=\boldsymbol{\Psi}_{k+1:p, k+1:p}$.
Their distribution can be derived by considering that each column $k$ results from the sum of the constant term $\psi_{jk}\cdot c_k=\frac{q_{kk}}{\psi_{kk}}\cdot \psi_{jk}$ and the linear combination of the independent standard Gaussian random variables $b_{jk},\,j=k+1,\dots,p$ according to coefficients $\psi_{jt},\,t=j+1,\dots, p$, i.e.: $    q_{jk}=\frac{q_{kk}}{\psi_{kk}}\cdot \psi_{jk}+\bPsi_{kk}\bB_{k+1:p,k}$.
This hierarchical representation of the distribution over $\bQ$, as described in Equations \eqref{eq:qdistdiag} and \eqref{eq:qdistoffdiag}, is probabilistically equivalent to the Wishart distribution. However, while the original Wishart representation involves a complex interdependence among all matrix elements, the one relying on the Bartlett decomposition introduces a structured representation in terms of conditionally independent components. This structure can be exploited to facilitate sampling \citep{pourahmadi1999joint} or computations involving Wishart-distributed random matrices and, in the case of the G-Wishart, has been used to design efficient algorithms for approximating the above-mentioned normalization constant \citep{atay2005monte}.

\section{A general hierarchical prior over sparse matrices}
\label{sec:hiersparse}
The G-Wishart distribution is based on the assumption that the density of random matrices in $\calP_p(\calG)$ is proportional to that of a standard Wishart distribution. While this assumption is reasonable, it presents the inconvenience of depending on an analytically intractable normalization constant. 
Here, we propose using the Cholesky factorization beyond its traditional role into.
The Cholesky decomposition allows factorizing any symmetric positive definite matrix $\bLambda$ into the product of a lower-triangular matrix $\bQ$ and its transpose. Conversely, every lower-triangular matrix $\bQ$ can be interpreted as the Cholesky factor of a symmetric positive definite matrix $\bLambda$. 
Interestingly, 
no specific mathematical constraints are required on the elements of $\bQ$ (other than $\lsq\bQ\rsq_{kk}>0,\,k=1,\dots,p$) for the product $\bQ\bQ^{\top}$ to be symmetric and positive-definite. 
Therefore, we can leverage this decomposition to define a distribution over the space of lower-triangular matrices $\calL_p$, say $f_\bQ(\cdot)$. Interpreting these matrices as Cholesky factors, we can map the space of all positive-definite matrices via $\bLambda=\mathbf{Q} \mathbf{Q}^{\top}$.
This also holds for sparse positive-definite matrices $\bLambda=\mathbf{Q} \mathbf{Q}^{\top}\in\calP_p(\calG)$, for which $\bQ\in\calL_p\lrnd\calG\rrnd$ (i.e. the space of lower triangular matrices such that the product by their transpose has a sparsity pattern corresponding to $\calG$). Importantly, unlike the direct manipulation of the elements of $\bLambda$, sparsity constraints can be effectively imposed by operating on the elements of $\bQ$ while still ensuring that the resulting $\bLambda$ remains positive definite.
Crucially, we can define a density $f_{\bQ|\calG}\lrnd\cdot\rrnd$ with support on $\calL\lrnd\calG\rrnd$ such that it integrates to $1$ naturally, eliminating the need for the numerical evaluation of any normalization constant.

From this point onward, the discussion is referred to a given graph $\calG=\lrnd\calV, \calE\rrnd$ even when its conditioning is not explicitly reflected in the notation.
Let us define a $p\times p$ symmetric binary matrix $\bZ$ associated to the graph  $\calG$ such that:
\begin{equation}
    \label{eq:zmat}
    z_{jk}=\lsq\bZ\rsq_{jk}=\lsq\bZ\rsq_{kj}=\begin{cases}
        &0\quad \text{if } (j,k)\notin \calE \text{ and } j>k\\
        &1\quad \text{if } (j,k)\in \calE \text{ or } j=k,
    \end{cases}
\end{equation}
with $j,k=1,\dots,p$.
The elements of $\bZ$ encode the sparsity pattern of the precision matrix $\bLambda$ by specifying which entries must be $0$ according to graph $\calG$, i.e.: $  z_{jk} = 0  \iff \lambda_{jk}=\lsq\bLambda\rsq_{jk}=\lsq\bLambda\rsq_{kj} = 0$.
The sparse precision matrix $\bLambda$ must have a corresponding Cholesky factorization $\mathbf{Q} \mathbf{Q}^{\top}$. 
The sparsity pattern of $\bLambda$ as determined by $\bZ$ is respected if and only if the elements of $\bQ$ corresponding to $z_{jk}=0$ satisfy specific relationships with the other elements of $\bQ$, while no constraints exist on the elements where $z_{jk}=1$. Specifically, for $1\leq k<j<p$:
\begin{equation}
\label{eq:sparsecondchol}
  \lambda_{jk} = 0 \iff q_{jk}=q^*_{jk}\lrnd\bQ_{1:k-1},q_{kk}\rrnd=
  \begin{cases}
  0 & \text{if } k = 1, \\
  -\frac{1}{q_{kk}}\cdot
  \sum_{t = 1}^{k-1} q_{jt}\cdot q_{kt}  & \text{otherwise,}
  \end{cases}
\end{equation}
where $\bQ_{1:k-1}$ denotes the first $k-1$ columns of $\bQ$. This provides a straightforward mechanism to set elements of $\boldsymbol{\Lambda}$ to $0$ by operating on the elements of $\bQ$. 
Note that Equation \eqref{eq:sparsecondchol} uniquely determines the $q_{jk}$ elements in the $k$-th columns such that $z_{jk}=0$ based on $\bQ_{1:k-1}$ and $q_{kk}$. 
Consequently, the diagonal elements can be sampled independently from a distribution $g_d(\cdot)$ with support on $\bbR^+$ and all off-diagonal elements can be sampled column-wise, in increasing order. In particular, those such that $z_{jk}=0$ must be set to $q^{*}(\bQ_{1:k-1}, q_{kk})$, while those such that $z_{jk}=1$ can be sampled from arbitrary distributions $g_o(\cdot)$ with support on $\bbR$.
This gives rise to the following conditional specification for the distribution over the elements of $\bQ$ given $\bZ$:
 \begin{equation}
 \label{eq:sparsedens}
     f_{\bQ\given \bZ}(\bQ)=\prod_{k=1}^pg_d(q_{kk})\cdot \prod_{k=1}^p\lsq\prod_{j:z_{jk}=1}g_o(q_{jk})\cdot\prod_{j:z_{jk}=0}\delta\lrnd q_{jk}\given q^*_{jk}\lrnd\bQ_{1:k-1}, q_{kk}\rrnd\rrnd\rsq,
 \end{equation}
where $q^*_{jk}\lrnd\cdot\rrnd$ is the mapping of Equation \eqref{eq:sparsecondchol} and $\delta(\cdot\given a)$ denotes the Dirac delta function in $a\in\bbR$. Notably, whatever the choice of $g_d(\cdot)$ and $g_o(\cdot)$, the density in Equation \eqref{eq:sparsedens} integrates to $1$ over the whole space of $\calL\lrnd\calG\rrnd$.
Algorithm 1 in Appendix A provides a pseudo-code for sampling a symmetric positive definite matrix with a sparsity pattern as determined by $\bZ$ through the hierarchical representation of Equation \eqref{eq:sparsedens}.

\subsection{The Sparse-Bartlett distribution}
\label{sec:Sbartlett}
The distributions over the non-zero elements of $\mathbf{Q}$ fully determine the distribution of the sparse precision matrix $\boldsymbol{\Lambda}$ for a given sparsity pattern $\bZ$. A natural choice for such distributions would be one coherent to the Bartlett decomposition introduced in Section \ref{sec:Sbartlett}. Indeed, such a choice allows recovering the original decomposition, hence a Wishart distributed precision matrix $\bLambda$, for a fully connected graph $\calG$ (i.e. $\bZ=\boldsymbol{1}_p$).

Following the Bartlett decomposition in Section \ref{subsec:bartlett}, we must first define the distribution over the diagonal elements $q_{kk},\,k=1,\dots,p$. The standard Bartlett decomposition of Equation \eqref{eq:qdistdiag} ascribes their squared values a Gamma distribution with the shape parameter numerator $\nu+p-k$, where $p-k$ is the number of non-zero entries below the diagonal in column $k$ of $\bLambda$. When inducing sparsity on $\bLambda$, some entries of $\bQ$ are set to given values to respect the zero constraints, resulting in the loss of degrees of freedom. This calls for an adjustment in the degrees of freedom of the corresponding diagonal element.
Therefore, we propose using the following distribution:
\begin{equation}
\label{eq:sbqdistdiag}
  q_{kk}^2 \sim Ga\left(\frac{\nu + z_k}{2}, \frac{1}{2\cdot \psi_{kk}^2}\right),
\end{equation}
where $z_k=\sum_{j=k+1}^{p}z_{jk}$ is the overall number of free entries in the $k$ column of $\bQ$, denoted from here on as $\bq_k$.

Conditionally on the diagonal elements, we can move to the definition of the distributions for the off-diagonal entries $q_{jk},\,1\leq k<j\leq p$. 
Let us define the set of non-zero elements below the diagonal in each column as $
    \calZ_k=\lcur j=k+1,\dots,p\,:\,z_{jk}=1\rcur
$,
with $k=2,\dots,p$ and size $|\calZ_k|=z_k$. At the same time, let $\bar{\calZ}_k$ be the set of zero elements below the diagonal of each column, with $|\bar{\calZ}_k|=\sum_{j=k+1}^p(1-z_{jk})$.
According to Equation \eqref{eq:qdistoffdiag} the columns are independent of one another. However, the off-diagonal elements in $q_{jk}\in\bar{\calZ}_k$ have a Dirac delta distribution on the function $q^*_{jk}(\cdot)$ of all previous columns. This inevitably induces marginal dependence also between the $q_{jk}\in\calZ_k$ belonging to different columns. This can be accounted for by means of the following conditional specification:
\begin{equation}
\label{eq:sbqdistoffdiag}
\begin{aligned}
    f_{\bq_{k}|\bZ, \bQ_{1:k-1}}&\lrnd\bq_{k}\rrnd=f_{\bq_{\bar{\calZ}_k, k}|\bZ, \bQ_{1:k-1}}\lrnd \bq_{\bar{\calZ}_k, k}\rrnd\cdot f_{\bq_{\calZ, k}|\bZ, q_{kk}, \bq_{\bar{\calZ}_k, k}}\lrnd\bq_{\calZ_k, k}\rrnd=\\
    &=\prod_{j\in\bar{\calZ_k}}\delta\lrnd q_{jk}\given q^*_{jk}\lrnd\bQ_{1:k-1}, q_{kk}\rrnd\rrnd\cdot
    \Norm_{z_k} \lrnd \mathbf{q}_{\calZ_k, k}\,\left| \, \bmu_c\lrnd q_{kk}, \bpsi_k, \bq_{\bar{\calZ}_k, k}\rrnd,\, \bSigma_c\lrnd\bPsi_k\rrnd\right.\rrnd,
    \end{aligned}
\end{equation}
where $\mu(\cdot)$ and $\bSigma(\cdot)$ are the conditional mean and variances resulting from the multivariate Gaussian in Equation \eqref{eq:qdistoffdiag}. Denoting as $\bphi$ and $\bV$ the marginal mean vector and variance-covariance matrix, we have:
\begin{equation*}
    \begin{aligned}
        &\bmu_c\lrnd q_{kk}, \bPsi_k, \bq_{\bar{\calZ}_k, k}\rrnd=\bphi_{\calZ_k}+\bV_{\calZ_k, \bar{\calZ}_k}\bV_{\bar{\calZ}_k, \bar{\calZ}_k}^{-1}\lrnd\bq_{\bar{\calZ}_k, k}-\bphi_{\bar{\calZ}_k}\rrnd,\\
        &\bSigma_c\lrnd \bPsi_k\rrnd=\bV_{\calZ_k, \calZ_k}-\bV_{\calZ_k, \bar{\calZ}_k}\bV_{\bar{\calZ}_k, \bar{\calZ}_k}^{-1}\bV_{\bar{\calZ}_k, \calZ_k}.
    \end{aligned}
\end{equation*}
Notably, this specification recovers a standard Wishart distributed matrix for a full $\bZ=\boldsymbol{1}_p$ and independent gamma distributed variances on the diagonal for a diagonal $\bZ=\IdentityMat_p$.

\section{Estimation via MCMC}
\label{sec:Estimation}
The proposed prior can be used as a prior for any positive definite matrix or covariance matrix. However, its most natural application is in the context of Bayesian Gaussian Graphical models, in which inference on the (potentially sparse) precision matrix $\bLambda$ of a multivariate Gaussian outcome is sought. In order to achieve that, we cannot rely on a given matrix $\bZ$ encoding the sparsity of $\bLambda$, but the elements of $\bZ$ must be explicitly modeled as random variables as follows:
\begin{equation}
\label{eq:priormod}
\begin{aligned}
    &\mathbf{Y}_i\given\boldsymbol{\Lambda}\, \sim\, \Norm_p(\bzero,\, \boldsymbol{\Lambda}^{-1}), \quad i = 1, \dots , n \\  
    &\boldsymbol{\Lambda}\given\bZ\, \sim\, \text{SB}(\nu,\, \mathbf{S},\, \bZ),\\
    &z_{jk} \sim Ber(\pi_{jk}), \quad k=1,\dots,p-1,\; j=k+1,\dots,p 
\end{aligned}    
\end{equation}
where $\text{SB}(\nu, \mathbf{S}, \bZ)$ denotes the newly introduced S-Bartlett and $\pi_{jk}$ represent the prior probability of $z_{jk}$ being different from $0$. 
Although the distribution $\text{SB}(\nu,\, \mathbf{S},\, \bZ)$ on $\bLambda$ does not present a captivating closed form expression, by recalling that $\bLambda=\bQ\bQ^{\top}$ we can directly adopt the following hierarchical specification:
\begin{equation} 
\label{eq:herarch_q}
    \begin{aligned}
     &\mathbf{Y}_i\given\bQ, \sim\, \Norm_p(\boldsymbol{\mu}_i,\, \bQ\bQ^{\top}), \quad i = 1, \dots , n\\
        & \mathbf{q}_{\calZ_k, k}\given \bQ_{\bar{\calZ}_k, k}, \bZ \sim \Norm_{z_k} \lrnd   \bmu_c\lrnd q_{kk}, \bPsi_k, \bq_{\bar{\calZ}_k, k}\rrnd,\, \bSigma_c\lrnd\bPsi_k\rrnd\rrnd,  \quad k = 1,\dots , p, \\
        &q_{jk}\given \bQ_{1:k-1}, \bZ\sim \delta\lrnd  q^*_{jk}\lrnd\bQ_{1:k-1}, q_{kk}\rrnd\rrnd, \quad k = 1,\dots , p, \quad j \in \bar{\calZ}_k,\\
        &q_{kk}^2 \given \bZ\sim Ga\left(\frac{\nu + z_k}{2}, \frac{1}{2\cdot \psi_{kk}^2}\right),  \quad k = 1,\dots , p,\\
        &z_{jk}=\lsq\bZ\rsq_{jk}\sim Ber(\pi_{jk}), \quad k=1,\dots,p-1,\; j=k+1,\dots,p 
    \end{aligned}
\end{equation}
Direct sampling of the elements of $\bQ$ is not trivial as they are in complex inter-relationships with one another, naturally calling for a joint update through an efficient Hamiltonian Monte Carlo (HMC) dynamic. However, mixed-type variables (continuous and discrete) are involved, with the number of free elements in $\bQ$ varying from iteration to iteration according to the current $\bZ$. In order to overcome all these complications without resorting to reversible jump MCMC routines, we opt for a transformational algorithm. Sampling is performed on the augmented space of a fully free lower triangular matrix $\bB$, which can then be linked to the desired sparsity-inducing version $\bQ$ through a deterministic transformation.
We sample the elements of the full $\bB$ from a typical Bartlett decomposition, with the only caveat of keeping into account the current number of non-zero elements as determined by $\bZ$ when sampling its diagonal elements. 
We will write $\bB\given \bZ\sim B(\nu+\bz,\,\IdentityMat_p)$ to denote:
\begin{equation*}
    \begin{aligned}
        &b_{kk}^2\given\bZ\,\sim\, Ga\lrnd\frac{\nu+z_k}{2},\,\frac{1}{2}\rrnd,\qquad b_{jk}\,\sim\,\Norm\lrnd 0, 1\rrnd,
    \end{aligned}
\end{equation*}
for $k=1,\dots,p,\; j=k+1,\dots,p$.
Given the sampled $\bB$, we can define the respective sparse version through a deterministic transformation $\bQ=Q^*(\bB)$, such that the resulting elements are distributed as in \eqref{eq:herarch_q}.
Exploiting the column-wise definition and the properties of multivariate Gaussian random variables, this can be achieved quite easily by defining: 
\begin{equation*}
\begin{aligned} 
\label{eq:trasf}
    & \mathbf{q}_{\calZ_k, k} =   \bmu_c\lrnd q_{kk}, \bPsi_k, \bq_{\bar{\calZ}_k, k}\rrnd +  \bSigma_c\lrnd\bPsi_k\rrnd^{\frac{1}{2}}\mathbf{b}_{\calZ_k, k} ,  \quad k = 1,\dots , p, \\
    &q_{jk}=  q^*_{jk}\lrnd\bQ_{1:k-1}, q_{kk}\rrnd, \quad k = 1,\dots , p, \quad j \in \bar{\calZ}_k,\\
    &q_{kk} = \psi_{kk}\cdot c_{k}, \quad k = 1,\dots , p.
\end{aligned}
\end{equation*}
It should be noted that the number of continuous random variables in the algorithm does not change based on the value of $\bZ$ as all elements of $\bB$ are sampled at all iterations. Note, however, that those in $\mathbf{b}_{\bar{\calZ}_k, k}$ do not contribute to the current $\bLambda$ and therefore to the posterior sample.
Efficiency in the sampling of the elements in $\bB$ is then achieved through a combination of No-U-Turn and Dual Averaging HMC \citep{nouturn}.

The sampling of $\bZ$ given all the other unknowns is quite standard. It follows the typical Stochastic Search Selection implementations \citep{mitchell1988bayesian}. Each $z_{jk}$ can only assume two values, hence its full conditional is $z_{jk}\given \cdots \sim Ber\lrnd\frac{p^1_{jk}}{p^1_{jk} + p^0_{jk}}\rrnd$ with:
\begin{equation*}
\begin{aligned}
    \label{eq:z}
&p^1_{jk} = \pi_{jk}\cdot Ga\lrnd{q_{kk}}^2\given \nu + z_k^1\rrnd\cdot \prod_{i=1}^n\Norm_p\lrnd \by_i\given \bzero,\, \bLambda^1\rrnd,\\
        &p^0_{jk} = (1-\pi_{jk})\cdot Ga\lrnd{q_{kk}}^2\given \nu + z_k^0\rrnd\cdot \prod_{i=1}^n\Norm_p\lrnd \by_i\given \bzero,\, \bLambda^0\rrnd,
\end{aligned}    
\end{equation*} 
where $\bLambda^1$ and $\bLambda^0$ represent the precision matrices computed assuming $z_{jk} = 1$ and $z_{jk} = 0$, respectively, and similarly for $z_j^1$ and $z_j^0$. 
Not that the updating probabilities in \eqref{eq:z} depend also on the marginal distribution of $b_{kk}^2$. Indeed, considering that their distribution depend on $z_k$, must take into account the change from an update in $\bZ$.


Most complex models involving additional parameters, such as a mean $\boldsymbol{\mu}_i=g(\boeta, \bx_i)$, can be easily integrated in the proposed MCMC routine by implementing ad-hoc Gibbs or Metropolis steps according to the specific application (e.g. linear regression or others) .

\subsection{GLMM Setting and Missing data}
\label{subsec:misglm}

Section \ref{sec:Estimation} is entirely dedicated to Gaussian Graphical Models. However, in many cases, the Gaussian assumption is irrealistic or incompatible with the observed data. When this is the case, the original setting can be easily extended to more general likelihoods in a Generalized Mixed Model framework by introducing latent Gaussian fields $\bW_i, i=1,\dots,n$ with distribution as specified in \eqref{eq:priormod}. The data distribution is then specified conditionally on this latent variables on top of the original hierarchy as:
\begin{equation*}
    \bY_i\given\bW_i\sim \prod_{j=1}^pf\lrnd \cdot\given g_{\mu}\lrnd W_{ij}\rrnd\rrnd,\quad i=1,\dots,n,
\end{equation*}
where $f(\cdot\given \mu)$ is a distribution with parameter $\mu$ linked to the latent variables through the function $g_\mu(\cdot)$.
If it belongs to the exponential family, then we can typically express such relationship on the natural parameter through a canonical link function, yielding:
\begin{equation}
g_{\mu}\lrnd\bbE\lsq Y_{ij}\given W_{ij}\rsq\rrnd = \mu_{j}+W_{ij},\quad j=1,\dots,p,
\end{equation}
where $\mu_{j}$ are outcome-specific intercepts.
This latent variables specification allows estimating the conditional independence structure for a broad family of distribution as long as posterior sampling for the latents $\bW_i$ is feasible. As a matter of fact, this can be achieved through an additional HMC step or, if the specific data likelihood allows, through ad-hoc Gibbs step updates.

This same latent variable scheme can be used to predict/impute missing data in the original Gaussian specification or the GLMM one. As a matter of fact, the model can be specified for the full outcome across all observations, which are partitioned into observed and missing data point as $\bY_i=\lsq\bY_i^o,\, \bY_i^m\rsq$. Then, at each iteration, the value of the missing data is updated conditionally on the current set of parameters and the other observed data. In the case of Gaussian Graphical Models, this requires sampling $\bY^m_{i}\given\mathbf{Y}^o, \boldsymbol{\Lambda}$ at each iteration, which can be easily derived thanks to the closed-form expression of the conditional distribution
of a multivariate normal. In the GLMM case, it just requires sampling $Y^m_{ij}\given W_{ij}^o, \boldsymbol{\Lambda},\; j=1,\dots,p$.

\section{Simulations}
\label{sec:sim}
We perform an extensive simulation study encompassing both the Gaussian and GLMM settings with three main objectives: (i) recovery of the true sparsity pattern, (ii) overall estimation of the underlying precision matrix $\bm{\Lambda}^*$, (iii) predictive performances for incomplete data. 
We consider outcomes of sizes $p\in \lbrace10, 25\rbrace$ and investigate two different sparsity patterns of the underlying $\bLambda^*$: banded matrices with bands of width $w\in\lbrace 1, 2, 3\rbrace$, named band$_w$; random sparsity pattern with sparsity proportions of $\alpha\in\lbrace0, 0.25, 0.5, 0.75\rbrace$, named rand$_\alpha$.
In the case of the banded matrices, we initially build a matrix $\bLambda$ with diagonal elements equal to 1 and non-zero off-diagonal elements to $-0.999/(2\cdot w)$ in order to ensure PD-ness, and then rescale it into $\bLambda^*$ to ensure that the corresponding $\bSigma^* = \lrnd\bLambda^*\rrnd^{-1}$ has diagonal elements all equal to $1$.
In the case of random sparsity matrices, these are generated from the G-Wishart prior as implemented in \cite{mohammadi2019bdgraph}. 
The expected values have been fixed to a constant $\mu_j =  \mu$, with $\mu=0$ for the Gaussian and $\mu = 5$ for the Poisson. 
For each scenario, we simulate $n=100$ observations for $N_{\text{sim}} = 20$ replicas.
The prior setting is is that of Equation \eqref{eq:priormod} with $\nu=3, \bS=\IdentityMat_p$ and all $\pi_{j,k}=0.5\;\forall\,j,k$. 
The No-U-turn algorithm is implemented  with parameter $M=M^{\text{adapt}}= 10$ and $\delta = 0.5$ \citep{nouturn}.
Whenever possible, the results are compared with those obtained with the G-Wishart weighted proposal algorithm (WWA) by \cite{Willem}, setting the prior parameters to $\beta = 3$ and $\bS=\IdentityMat_p$. Both MCMC algorithms are run for 10,000 iterations, with a burn-in of 8,000.
To summarise the results,  we report the sensitivity in the recovery of the true zero elements in $\bm{\Lambda}$. In addition, we report the estimated number of non-zero edges and the KL-discrepancy between $\hat{\bm{\Lambda}}$ and $\bm{\Lambda}$ \citep{raymaekers2023cellwise} to measure how well the true precision is recovered as a whole. Finally, we use the Continuous Rank Probability Score (CRPS) by \cite{Gneiting2007} to assess the predictive ability. For the sake of brevity, we only report the results related to the Normal likelihood herein and illustrate the ones for the Poisson case in Appendix C. 

\subsection{Simulation results: Normal}
\begin{figure}[t]
    \centering
    \begin{subfigure}[b]{.49\textwidth}
    \centering
    \includegraphics[width=0.99\linewidth]{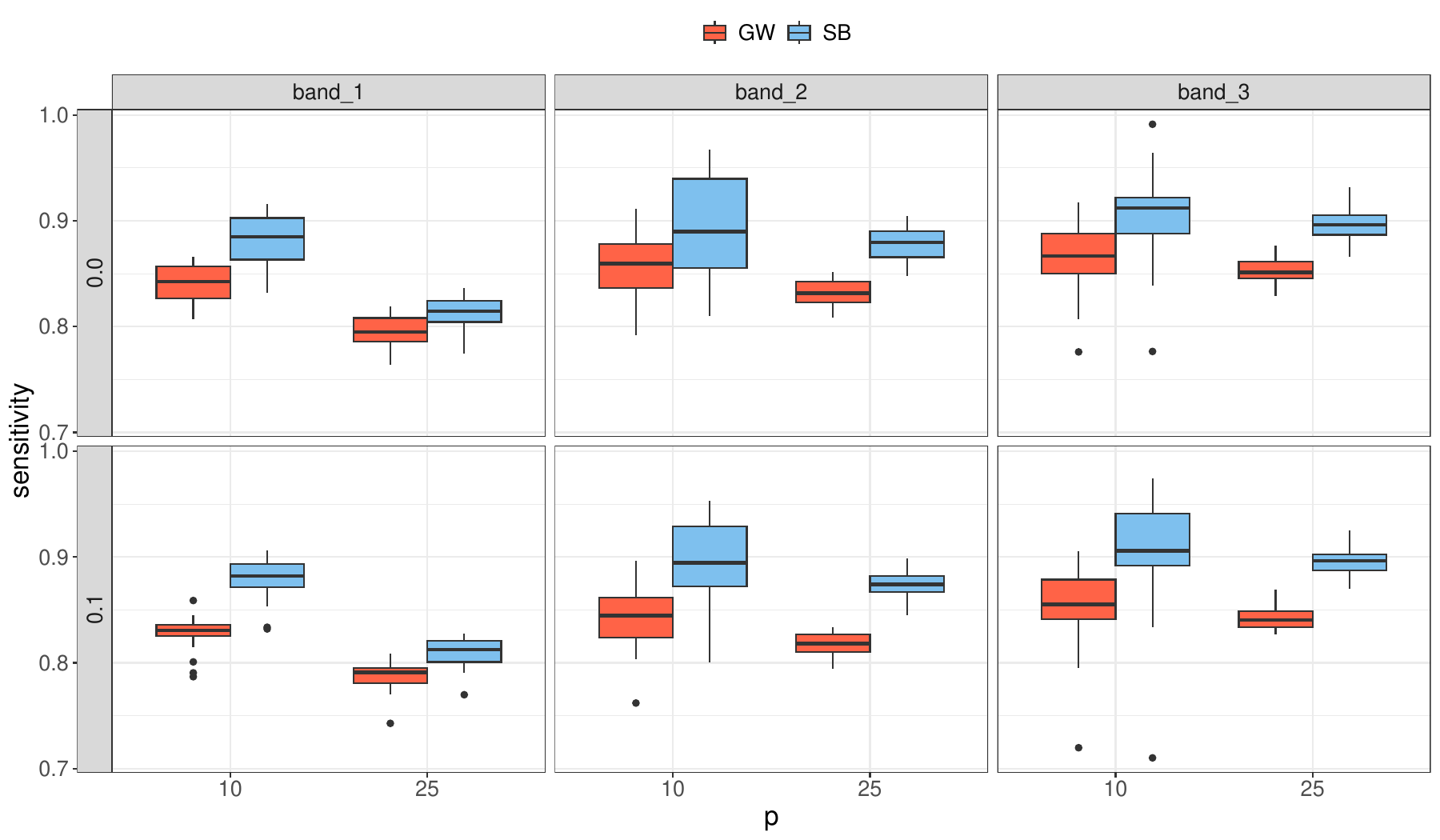}
    \caption{}
    \label{fig:sim_compsens_banded}
    \end{subfigure}
    \begin{subfigure}[b]{.49\textwidth}
    \centering
        \includegraphics[width=0.99\linewidth]{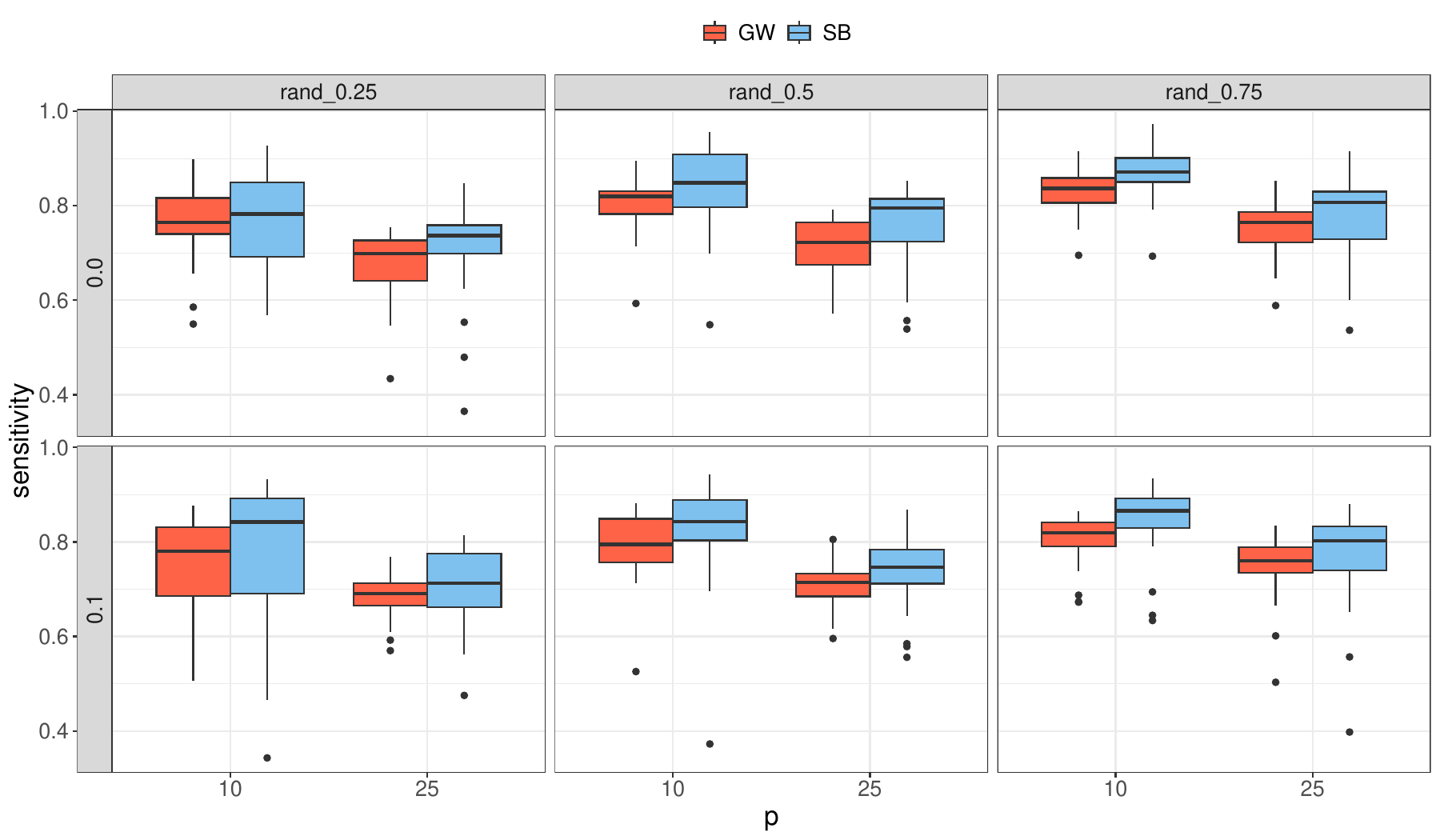}
        \caption{}
        \label{fig:sim_compsens}
    \end{subfigure}
    \caption{Sensitivity of the zero recovery of the G-Wishart and the S-Bartlett by varying the missingness (rows) and the sparsity proportion (cols) for banded (a) and random (b) sparsity patterns.}
\end{figure}

\begin{figure}[t]
    \centering
    \begin{subfigure}[b]{.49\textwidth}
     \centering
     \includegraphics[width=0.99\linewidth]{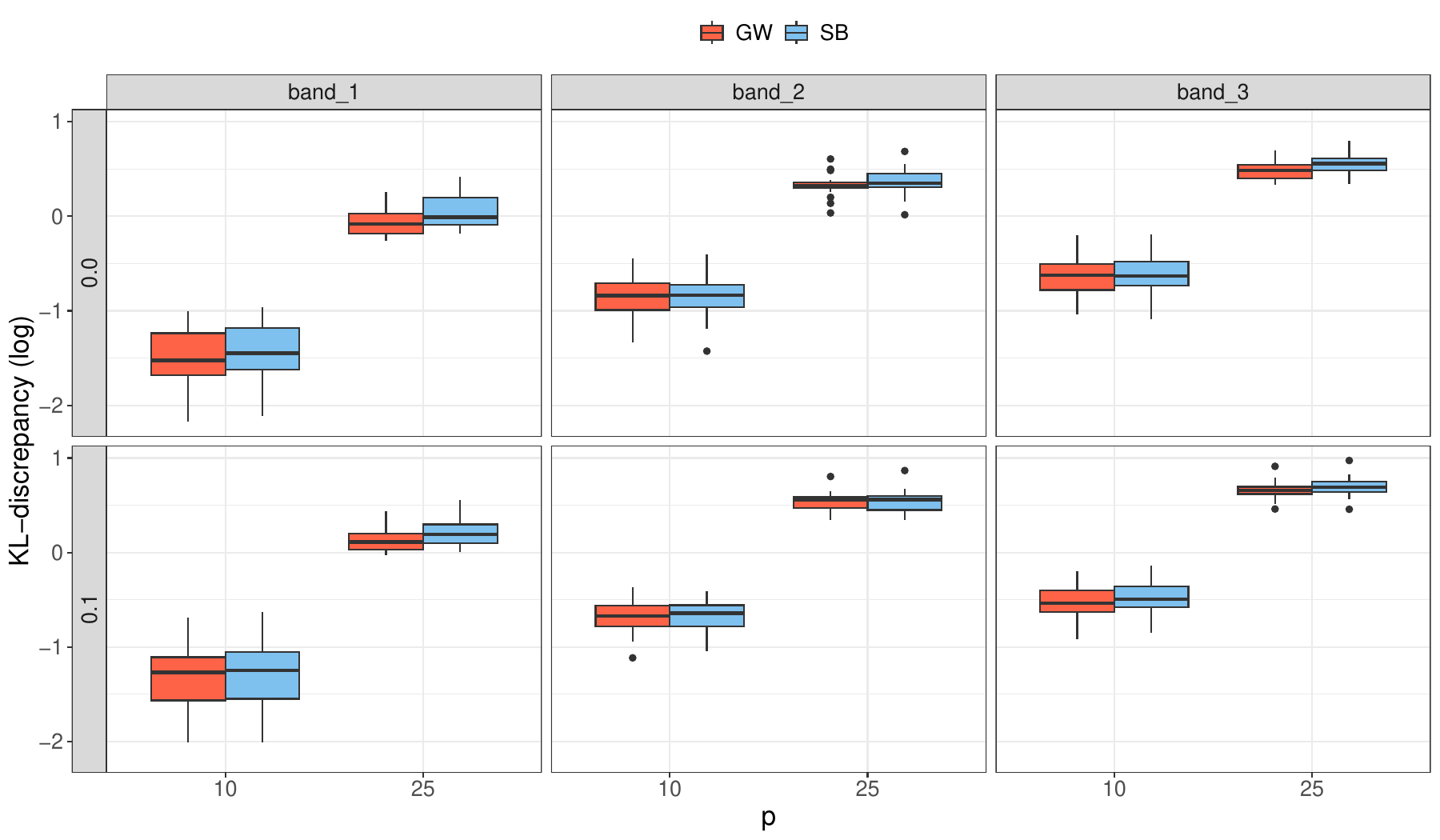}
    \caption{}
     \label{fig:sim_compKL_banded}
    \end{subfigure}
      \begin{subfigure}[b]{.49\textwidth}
      \centering
       \includegraphics[width=0.99\linewidth]{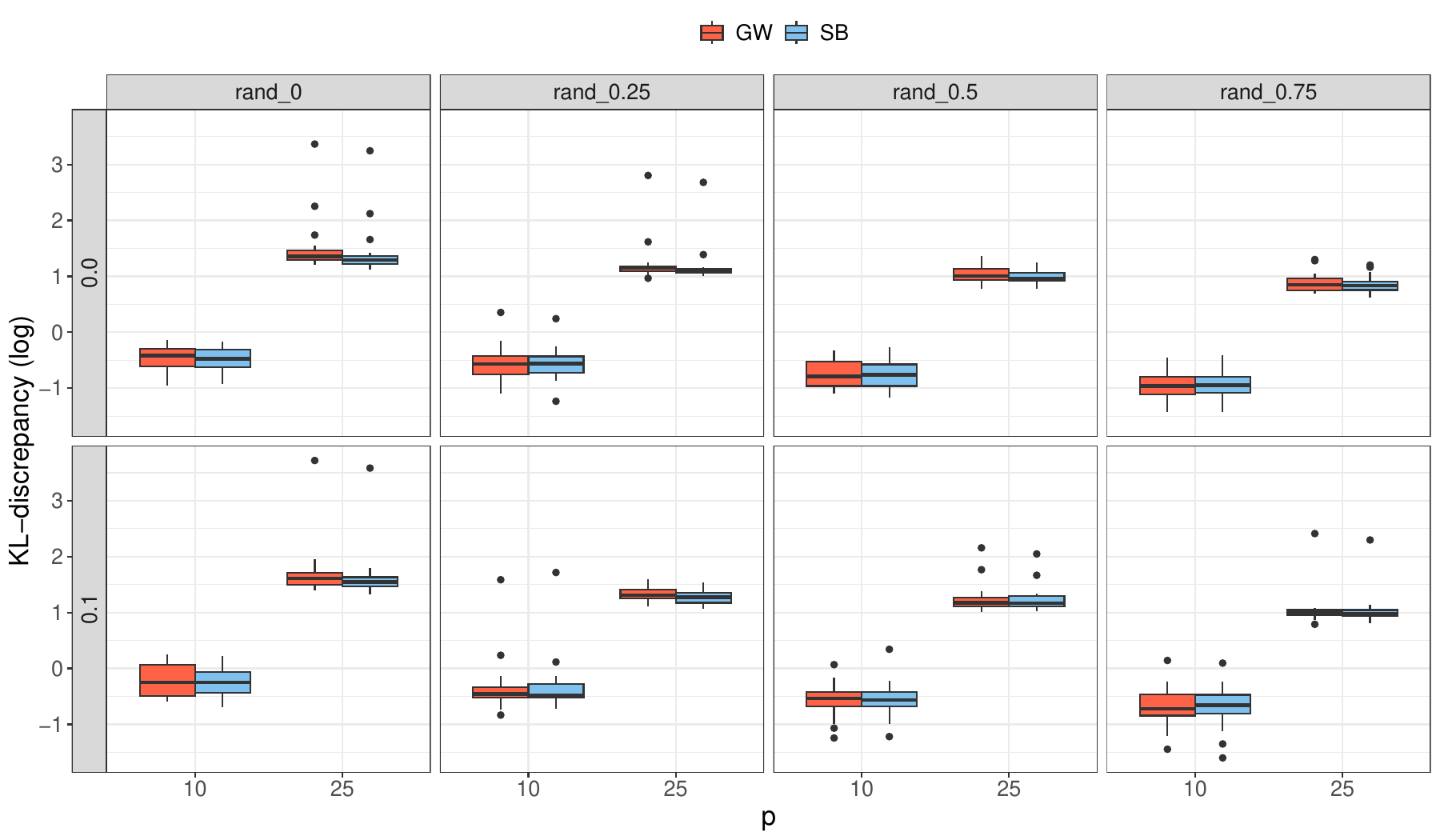}
        \caption{}
         \label{fig:sim_compKL}
    \end{subfigure}
    \caption{KL-discrepancy between the true $\bLambda$ and estimated $\hat{\bLambda}$ by the G-Wishart and the S-Bartlett by varying the missingness (rows) and the sparsity proportion (cols) for banded (a) and random (b) sparsity patterns.}
\end{figure}

First of all, we compare the sensitivity (percentage of \textit{correctly classified} zeroes) 
of the S-Bartlett and the G-Wishart for varying outcome's dimension and missingness percentages. Note that the final sparsity pattern is estimated by averaging $\bZ^{(\ell)}, \ell=1,\dots,L$ across all iteration into $\hat{z}_{jk}=\frac{1}{L}z_{jk}^{(\ell)}$ and then setting to $0$ those entries such that $\hat{z}_{jk}<0.5$. Figures \ref{fig:sim_compsens_banded} and \ref{fig:sim_compsens} show the results for the banded and random sparsity, respectively.
The results consistently favor the S-Bartlett proposal, suggesting its improved ability in the recovery of the true sparsity pattern of $\bLambda$. 
Figures \ref{fig:sim_compKL_banded} and \ref{fig:sim_compKL} compare the KL-discrepancy between the \textit{true} $\bLambda$ and the final Bayesian estimator derived by the two competitors. While in the banded setting the G-Wishart prior appears to offer a slightly more accurate estimate of $\bLambda$, the S-Bartlett provides slightly better performances (i.e. smaller KL-discrepancies) in all scenarios with random sparsity patterns.
Finally, Tables \ref{tab:compare_crpsnormalrandom} and \ref{tab:compare_crpsbanded} report the CRPSs of the missing-data predictions. We note that the S-Bartlett returns smaller errors, exhibiting improved performances in the prediction of missing data. These improved scores are consistently accompanied by a lower number of non-zero entries in $\bLambda^{(\ell)}$ across all iterations, as measured by $\bar{z}^{(\ell)} = \sum_{jk} z_{jk}^{(\ell)}$ (see Appendix B). This highlights the greater parsimony of the S-Bartlett approach, achieved without any loss of information. 
\begin{table}
    \caption{Average CRPS obtained by the G-Wishart and the proposed S-Bartlett for banded sparsity patterns.}
    \centering
    \begin{tabular}{ll|ccc|ccc}
        $pmiss=10\%$ &  & \multicolumn{3}{c}{G-Wishart} & \multicolumn{3}{c}{S-Bartlett}\\
         \midrule
         Banded & & 1 & 2 & 3 & 1 & 2 & 3 \\
          \cline{3-8}
        \multirow{2}{*}{$k$} & $10$ & 0.306 & 0.410 & 0.470 & 0.305 & 0.407 & 0.469 \\
         & $25$ & 0.229 & 0.320 & 0.386 & 0.228 &  0.318 & 0.386 \\
         \bottomrule
    \end{tabular} \label{tab:compare_crpsbanded}
\end{table}
\begin{table}
    \caption{Average CRPS obtained by the G-Wishart and the proposed S-Bartlett for random sparsity patterns.}

    \centering
    \begin{tabular}{ll|cccc|cccc}
         $pmiss=10\%$ &  & \multicolumn{4}{c}{G-Wishart} & \multicolumn{4}{c}{S-Bartlett}\\
         \midrule
        \% Sparsity  & & 0.0 & 0.25 & 0.5 & 0.75 & 0.0 & 0.25 & 0.5 & 0.75 \\
          \cline{3-10}
        \multirow{2}{*}{$k=$} & $10$ & 0.298 & 0.339 & 0.379 & 0.372 & 0.296 & 0.337 & 0.378 & 0.371 \\
         & $25$ & 0.224 & 0.244 & 0.268 & 0.285 & 0.221 & 0.240 & 0.265 & 0.284\\
         \bottomrule
    \end{tabular}    \label{tab:compare_crpsnormalrandom}
\end{table}

\section{Real data application} \label{sec:application}
In this section, we illustrate two real data applications. For the first one, we assume the data come from a Gaussian distribution; for the second one, we assume the data come from a Poisson distribution. In the former case, a comparison is made with the G-Wishart to test possible differences with the S-Bartlett proposal, expanding beyond the simulation study. For each dataset, the number of iterations, burnin, thinning, and the No-U-Turn sampler parameters are the same as in the simulation study of Section \ref{sec:sim}.

\subsection{Gene expression data}\label{subsec:genes}
We consider the real data application from Section 4.2 of \cite{mohammadi2015bayesian} that has been used to test the proposal by \cite{Willem}. We use the latter as a benchmark to compare the performance of the proposed S-Bartlett. The data consist of gene expressions in B-lymphocyte cells from $n = 60$ individuals that have been quantile-normalized to marginally follow a standard Gaussian distribution. We consider two datasets consisting of the first $p = 50$ and $p = 100$ most variable gene expressions. For both datasets, we randomly remove $10\%$ of the observations and treat them as additional parameters to be recovered within the Bayesian estimation routine. 

For $p=50$ and for $p=100$, the percentages of estimated zero entries are 92.4\% and 90\% for the SB, and 94.2\% and 90\% for the GW, respectively. This shows quite similar results across the two methods. However, in terms of the number of non-zero entries across each single iteration, the SB consistently stations on a lower value. In particular, for $p=50$ the average value of $\bar{z}^{(\ell)}$ is $172 \,(155, 191)$ for the SB and $215 \,(193, 235)$ for the GW. The same pattern is visible when the dimensionality rises to $p=100$, with SB averaging $846 \, (801, 893)$ non-zeros against the GW’s $1022 \, (968, 1078)$. This corresponds to a consistent surplus of $\approx$ 21\% of the edges by the GW.
This difference in the two methods is due to the certainty with which each entry is set to $0$ by the two methods. Appendix D reports details on the distributions of all $\hat{z}_{jk}, 1\leq j<k\leq p$, highlighting how those of the SB are polarized toward $0$ or $1$ (i.e. more certain), while the GW presents more uncertain results (especially for the zero elements). 
In terms of predictive performances, as measured by the CRPS averaged across missing observations, SB outperforms GW in both settings. For $p=50$, the median CRPS is 0.767 for SB and 0.794 for GW; for $p=100$, the values are 0.697 and 0.721, respectively. Since lower CRPS values indicate more accurate probabilistic forecasts, this supports how the SB achieves better predictive accuracy than GW across both scenarios.
\begin{figure}
    \centering
    \begin{subfigure}[b]{.48\textwidth}
    \centering
    \includegraphics[width=.95\textwidth]{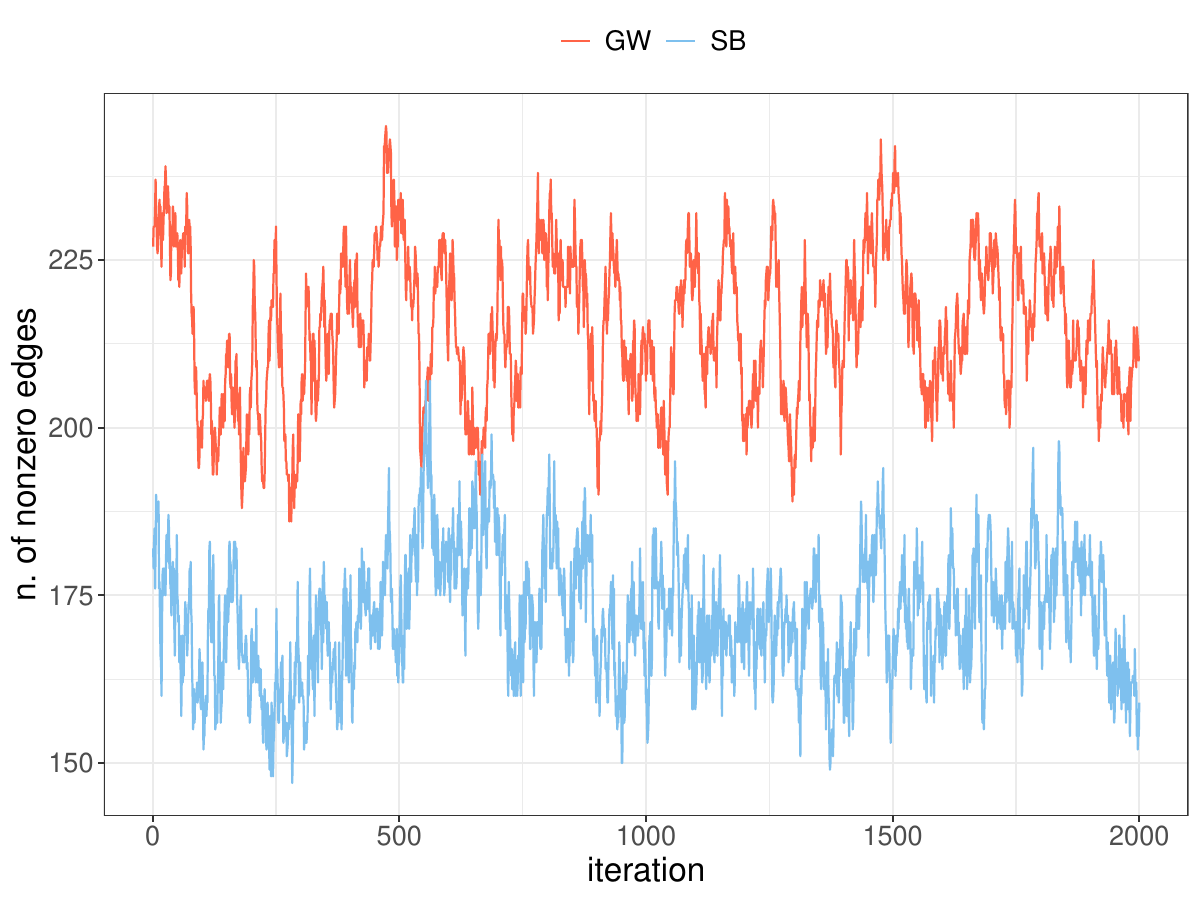}
    \caption{}
    \label{fig:trace_gene50}
    \end{subfigure}
    \begin{subfigure}[b]{.48\textwidth}
    \centering
    \includegraphics[width=.95\textwidth]{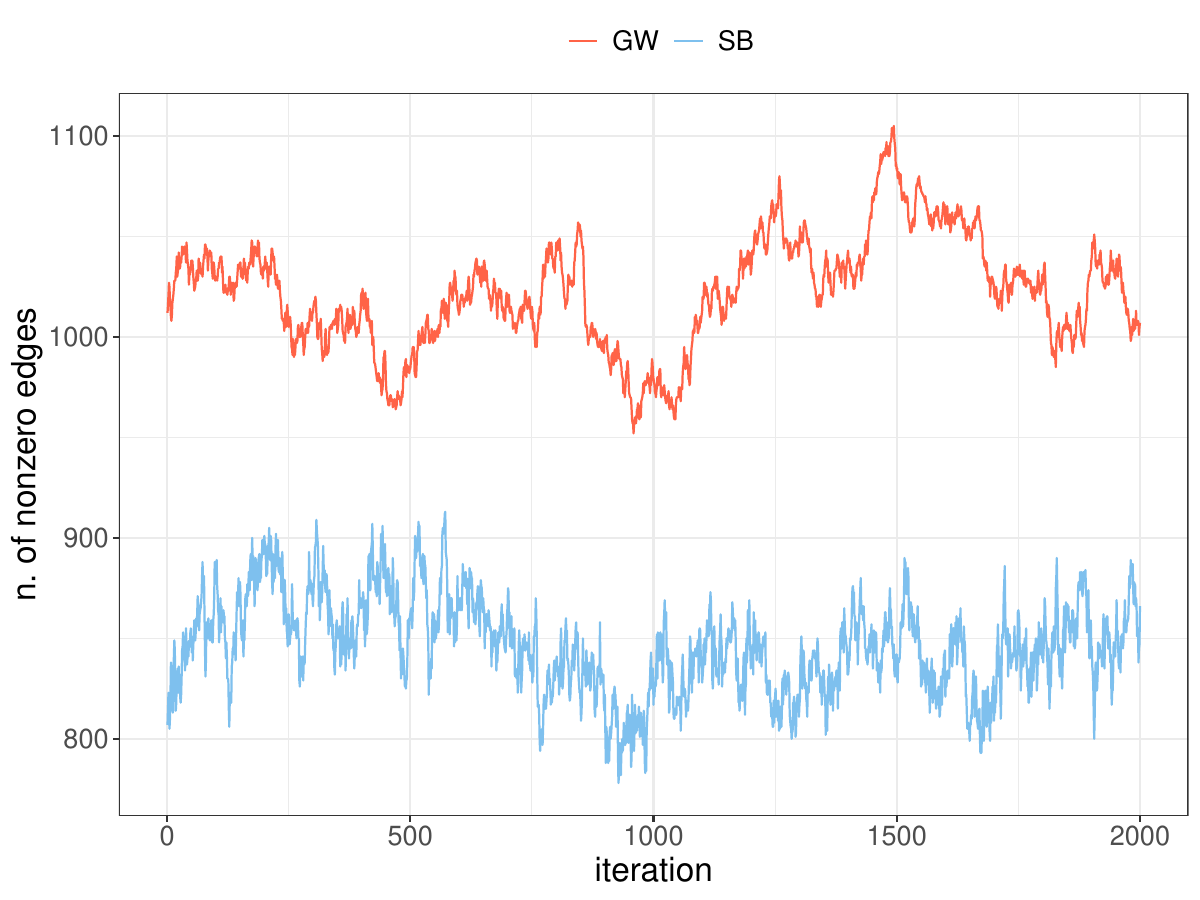}
    \caption{}
    \label{fig:trace_gene100}
    \end{subfigure}
    \caption{Traceplots of the number of non-zero edges estimated by the G-Wishart and the S-Bartlett for the gene expression dataset with $p = 50$ (a) and (b) $p = 100$.}
    \label{fig:trace_gene}
\end{figure}

\subsection{Doubs data}\label{subsec:doubs}
\begin{figure}
    \centering
    \begin{subfigure}[b]{.48\textwidth}
    \centering
    \includegraphics[width=.9\textwidth]{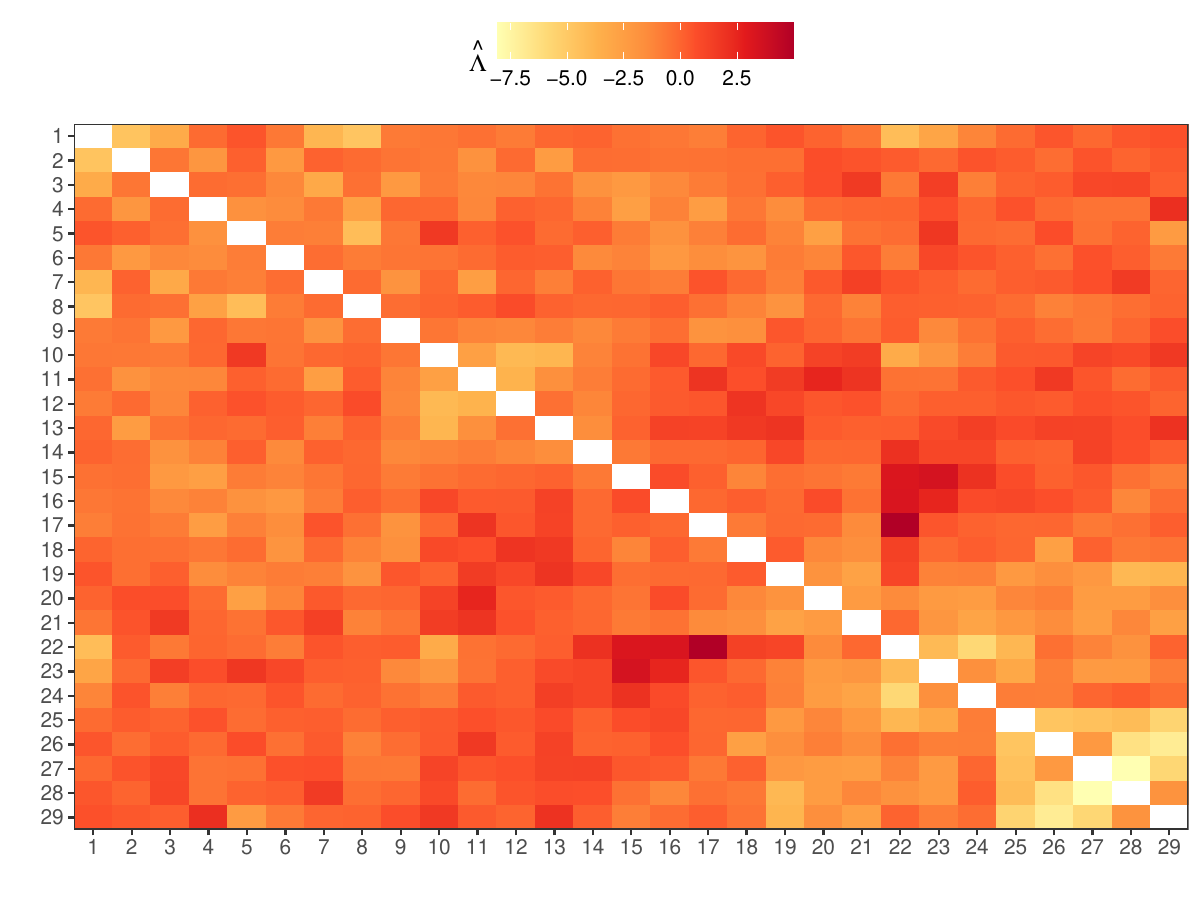}
    \caption{}
    \label{fig:estL_pois}
    \end{subfigure}
    \begin{subfigure}[b]{.48\textwidth}
    \centering
    \includegraphics[width=.9\textwidth]{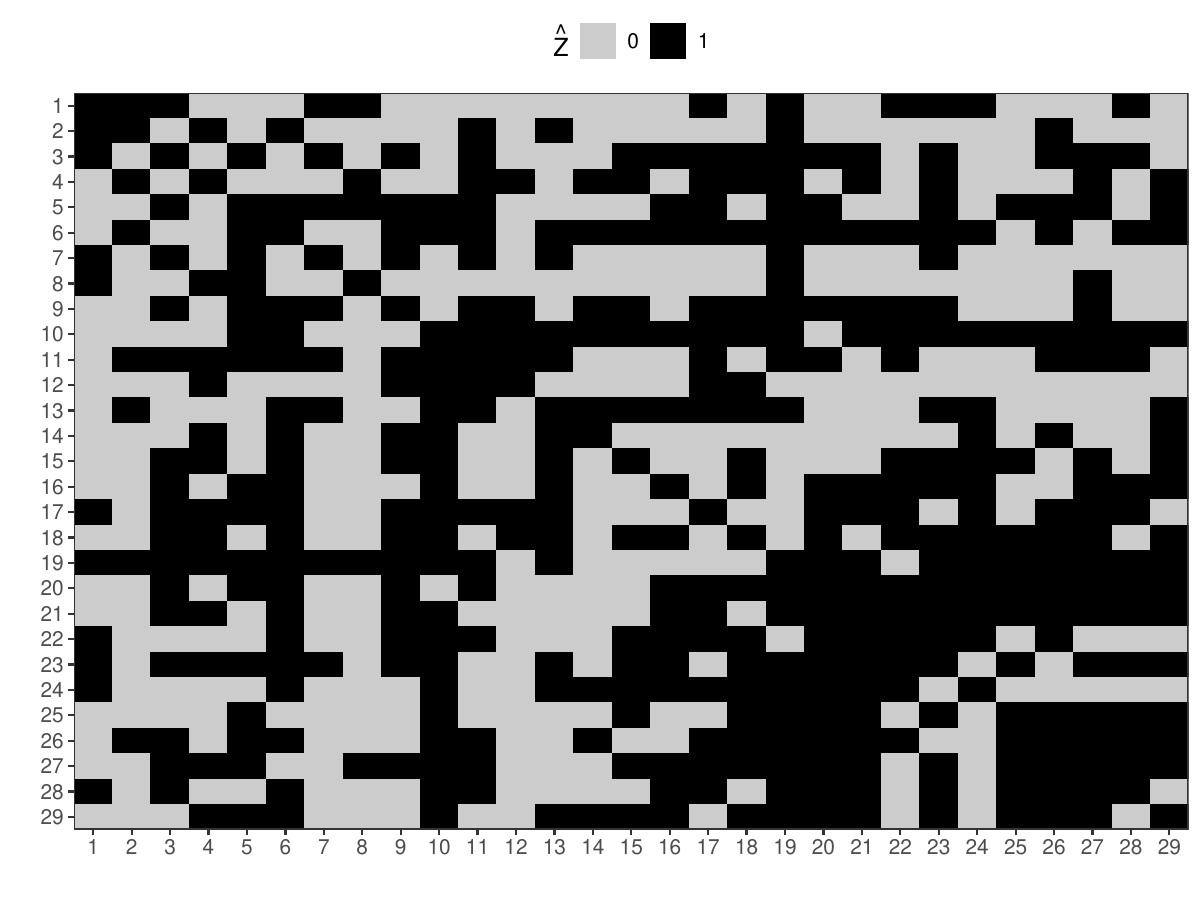}
    \caption{} 
    \label{fig:estZ_pois}
    \end{subfigure}
    \caption{Estimated $\hat{\bLambda}$ (a) and $\hat{\bZ}$ (b) for doubs data.}
    \label{fig:enter-label}
\end{figure}

For the second application, we use a well-known dataset on fish catches collected along the Doubs River, which runs near the France-Switzerland border in the Jura Mountains \citep{borcard2011numerical}. The data includes counts ($>0$) of fish species collected over 29 sampling hauls. The counts go from 0 to 5, with an average of 1, and the most abundant site is n. 29, counting a total of 89 collected fish across all species. The goal is to estimate the dependence structure among these hauls using abundance data in the proposed GLMM setting with the SB prior. 
The estimated $\bLambda$ is reported in Figure \ref{fig:estL_pois} and the related $\hat{\bZ}$ is shown instead in Figure \ref{fig:estZ_pois}. 
The results highlight a dense pattern, which seems to capture unobserved heterogeneity driven by spatial proximity as a proxy for environmental features shared by the sampled hauls. In particular, an overall sparsity of $\approx 48.5\%$ is estimated, with an expected number of non-zero edges of $203 \, (186, 219)$.

\section{Conclusive remarks}\label{sec:conclusion}
In this work, we introduced a novel hierarchical prior, the S-Bartlett, which encompasses the Wishart and diagonal independent distributions as particular cases. The method is based on the Cholesky factorization of the precision matrix and imposes sparsity via direct constraints on the elements of the Cholesky factors while preserving their conditional independence. Unlike previous approaches, this prior circumvents the need to compute the typically intractable normalizing constant. This construction not only enhances computational tractability but also offers greater flexibility in prior specification. Indeed, the proposed framework and its estimation routine can be naturally embedded into more general frameworks (e.g. GLMM settings), enabling the modeling of complex dependencies in multivariate count data or other types of non-continuous outcomes.

Our empirical evaluations show that the proposed method performs particularly well in identifying the true sparsity pattern of the underlying graphical model, regardless of its structural complexity. When it comes to handling missing data, performances proved to do better better than existing methods both on simulated and real data. In general, the S-Bartlett seems to favor more parsimonious models without sacrificing relevant information, which may be advantageous in practical applications where model simplicity and interpretability are desirable. 



\paragraph*{Acknowledgements}
This work has been supported by MIUR, grant number 2022XRHT8R - The SMILE project: Statistical Modelling and Inference for Living the Environment.\\ P. Alaimo Di Loro has been partially supported by PON “Ricerca e Innovazione” 2014-2020 (PON R\&I FSE-REACT EU), Azione IV.6 “Contratti di ricerca su tematiche Green”, grant number 60-G-34690-1.

\paragraph*{Declaration of interest}
The authors report there are no competing interests to declare.

\paragraph*{Code and data availability}
Data and codes to reproduce the results from the simulation study and the applications are available at https://github.com/minmar94/sparseBartlett.


\newpage
\appendix

\section{Details of the sampling scheme}

Algorithm \ref{alg:SparseSample} provides a schematic overview of the sampling scheme used to sample a sparse symmetric precision matrix as defined in Equation (8).

\begin{algorithm}[H]
\small
	\KwIn{$p$, $\calG$, $L$, $g_o(\cdot)$, $g_d(\cdot)$. Obtain $\bZ$ from $\calG$.}
	\For{$\ell=1,\dots,L$}{
        \For{$j=1,\dots,p$}{
            Sample $q_{jj}\sim g_d(\cdot)$\\
            \If{$j>1$}{ 
            \uIf{$z_{j1}=1$}{
                sample $q_{j1}\sim g_o(\cdot)$}
            \Else{set $q_{j1}=0$}
    }}
    \For{$k=2,\dots,p$}{
        \For{$j=2,\dots, k-1$}{
            \uIf{$z_{jk}=1$}{
                sample $q_{jk}\sim g_o(\cdot)$}
            \Else{set $q_{jk}=-\frac{\sum_{t = 1}^{k-1} q_{jt}\cdot q_{kt}}{q_{kk}}$}
    }}}
    Obtain $\bLambda^{(\ell)}=\bQ\bQ^{\top}$\\
	\KwOut{sample of symmetric positive definite matrices $\bLambda^{(\ell)}, \ell=1,\dots,L$}
	\caption{Sampling algorithm for sparse symmetric precision matrices}
\label{alg:SparseSample}
\end{algorithm}
\newpage
\section{Simulation results: Normal}

\begin{figure}[H]
    \centering
     \begin{subfigure}[b]{.48\textwidth}
     \centering
       \includegraphics[width=0.99\linewidth]{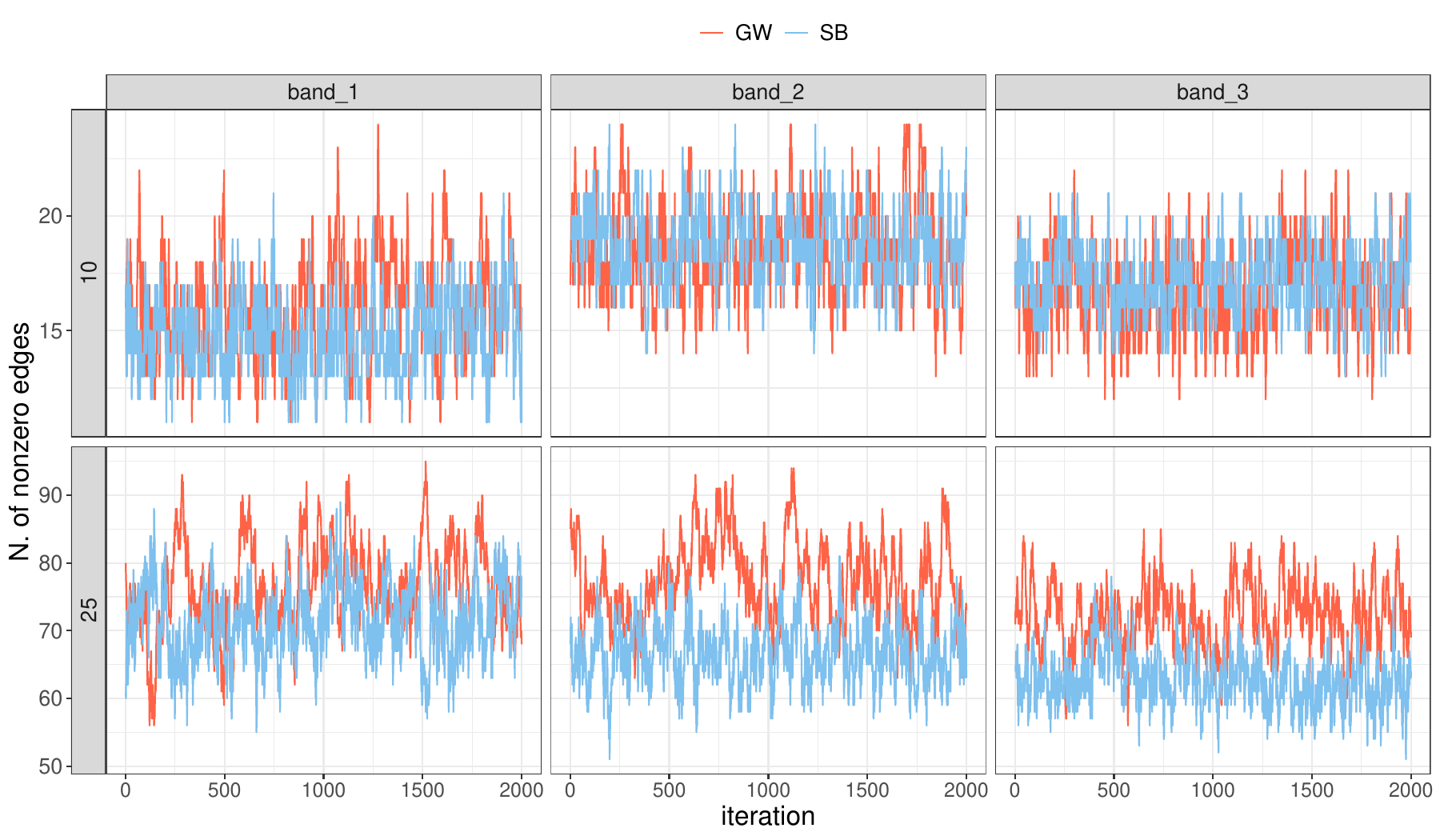}
        \caption{}
          \label{fig:tracecompare_nzerobanded}
    \end{subfigure}
    \begin{subfigure}[b]{.48\textwidth}
    \centering
        \includegraphics[width=0.99\linewidth]{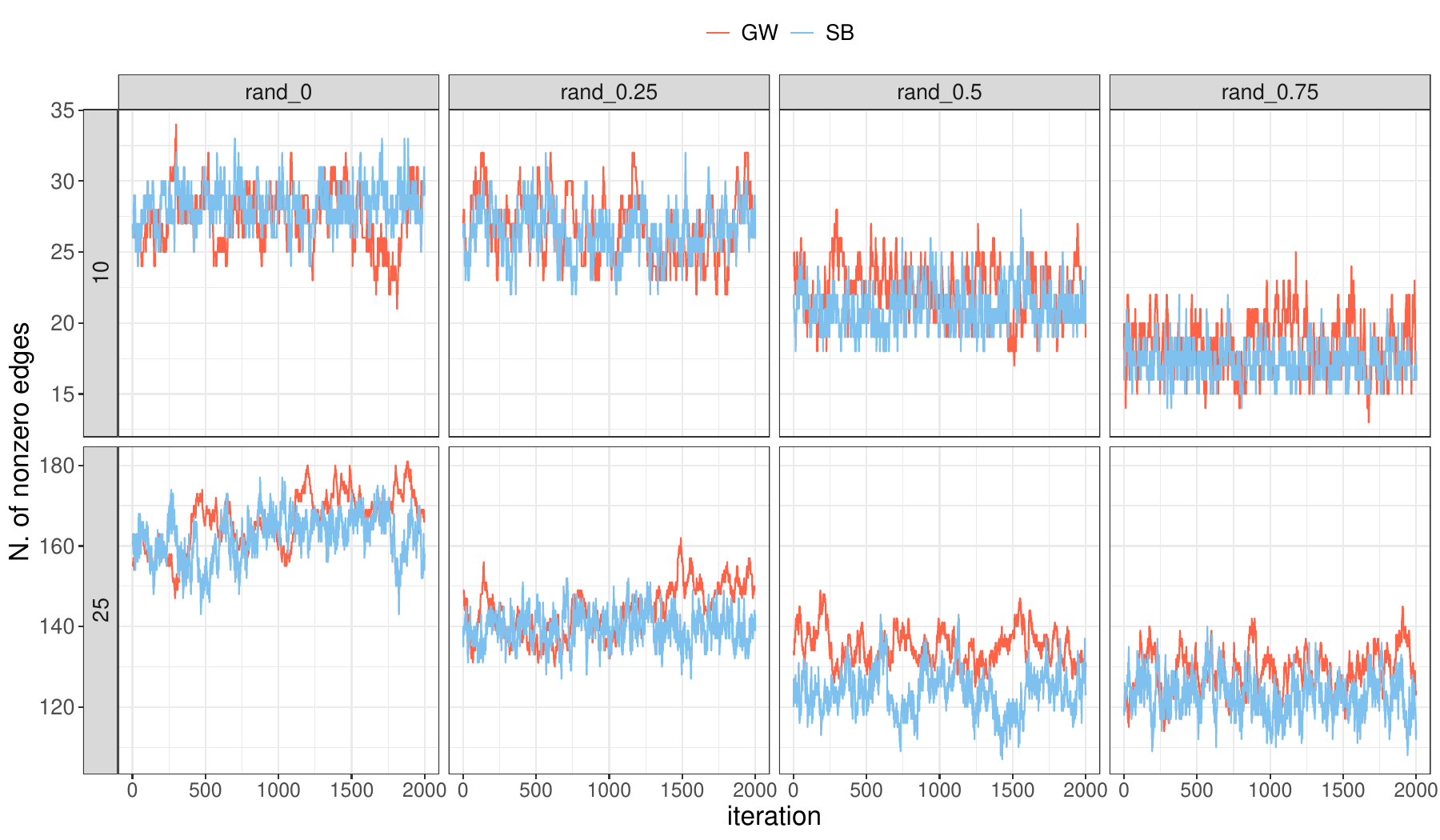}
        \caption{}
         \label{fig:tracecompare_nzerorandom}
    \end{subfigure}
    \caption{Traceplots of the number of nonzero edges estimated by the G-Wishart and the S-Bartlett for one simulated dataset by varying the outcome dimension (rows) and the sparsity proportion (cols) for banded (a) and random (b) sparsity patterns with no missing data.} 
\end{figure}

Figure \ref{fig:tracecompare_nzerobanded} and Figure \ref{fig:tracecompare_nzerorandom} show the traceplots of the number of non-zero edges obtained at each iteration by the G-Wishart and the S-Bartlett on one simulated replica by varying the outcome dimension and the sparsity type with no missing data. The figures highlight that, on average, the number of non-zero edges is always lower for the S-Bartlett both for the banded structure (see Table \ref{tab:compare_nedgesbanded}) and the random sparsity pattern (see Table \ref{tab:compare_nedgenormalrandom}).

\begin{table}[H]
    \caption{Mean (95\% CI) number of non-zero edges obtained by the G-Wishart and the proposed S-Bartlett for banded sparsity patterns and no missing data.}
    \centering
     \adjustbox{max width=\textwidth}{%
    \begin{tabular}{ll|ccc|ccc}
         &  & \multicolumn{3}{c}{G-Wishart} & \multicolumn{3}{c}{S-Bartlett}\\
         \midrule
         Banded & & 1 & 2 & 3 & 1 & 2 & 3 \\
          \cline{3-8}
        \multirow{2}{*}{$k$} & $10$ & 16 (12, 20) & 19 (15, 23) & 17 (13, 20) & 15 (12, 18) & 19 (16, 22) & 17 (14, 20) \\
         & $25$ & 76 (63, 89) & 77 (67, 89) & 72 (62, 81) & 71 (61, 82) & 66 (59, 76) & 63 (56, 71) \\
         \bottomrule
    \end{tabular} 
    }
    \label{tab:compare_nedgesbanded}
\end{table}
\begin{table}[H]
    \caption{Mean (95\% CI) number of non-zero edges obtained by the G-Wishart and the proposed S-Bartlett for random sparsity patterns and no missing data.}

    \centering
    \adjustbox{max width=\textwidth}{%
    \begin{tabular}{ll|cccc|cccc}
         &  & \multicolumn{4}{c}{G-Wishart} & \multicolumn{4}{c}{S-Bartlett}\\
         \midrule
        \% Sparsity  & & 0.0 & 0.25 & 0.5 & 0.75 & 0.0 & 0.25 & 0.5 & 0.75 \\
          \cline{3-10}
        \multirow{2}{*}{$k=$} & $10$ & 27 (23, 31) &   27 (23, 32)  &  22 (19, 26)  &  18 (15, 22)    & 28 (25, 31) &   26 (23, 30)  &  21 (19, 24) &   17 (15, 20)  \\
         & $25$ & 166 (152, 178) & 144 (133, 156) & 135 (127, 144) & 129 (119, 139) & 162 (150, 172) & 140 (132, 148) & 124 (113, 135) & 123 (113, 133)\\
         \bottomrule
    \end{tabular}   
    }
    \label{tab:compare_nedgenormalrandom}
\end{table}

\newpage
\section{Simulation results: Poisson}
\label{appsec:simpois}

\begin{figure}[H]
    \centering
      \begin{subfigure}[b]{.48\textwidth}
        \centering
        \includegraphics[width=0.99\linewidth]{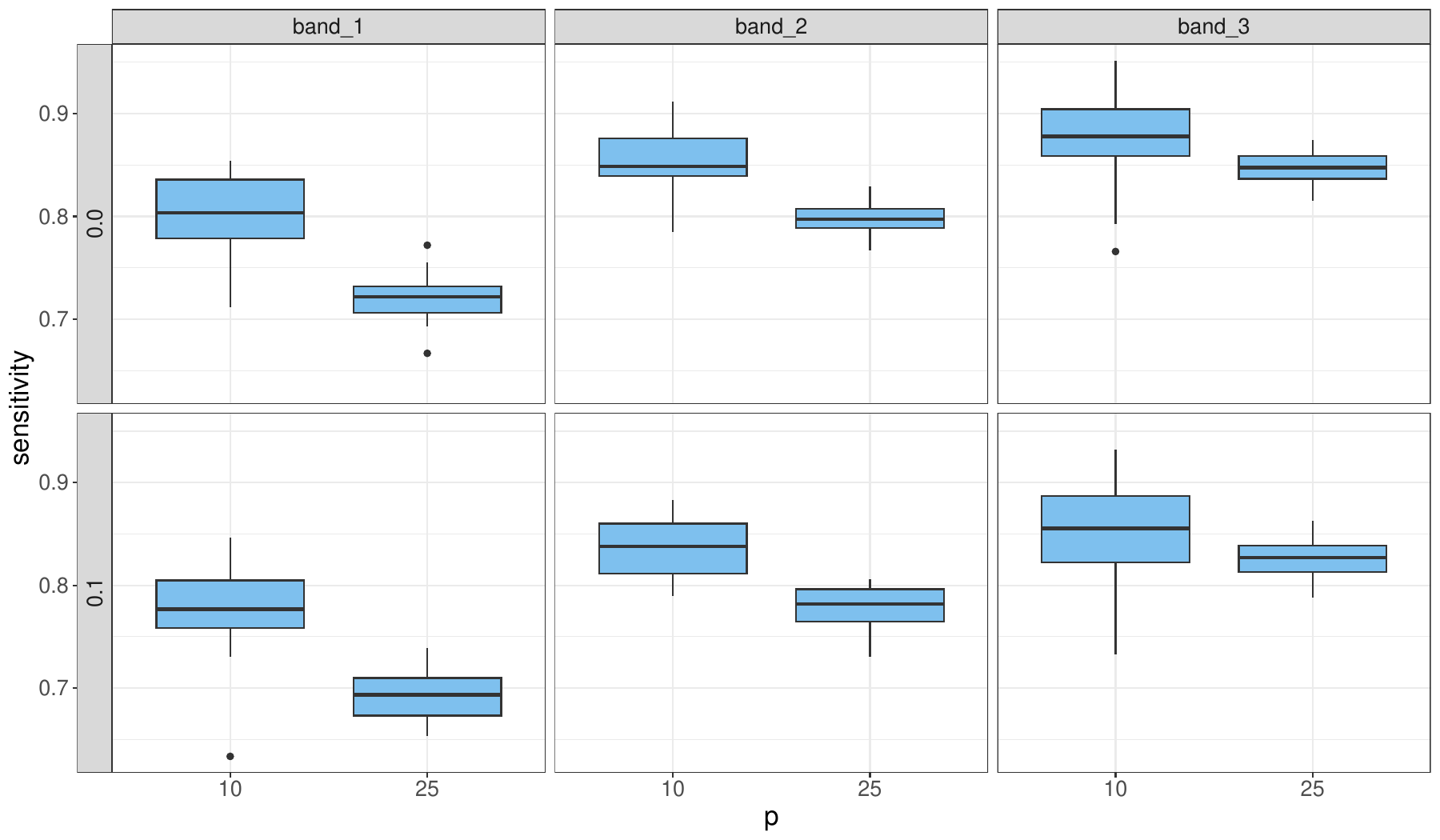}
        \caption{}
        \label{fig:sim_compsens_banded_pois}
    \end{subfigure}
    \begin{subfigure}[b]{.48\textwidth}
        \centering
        \includegraphics[width=0.99\linewidth]{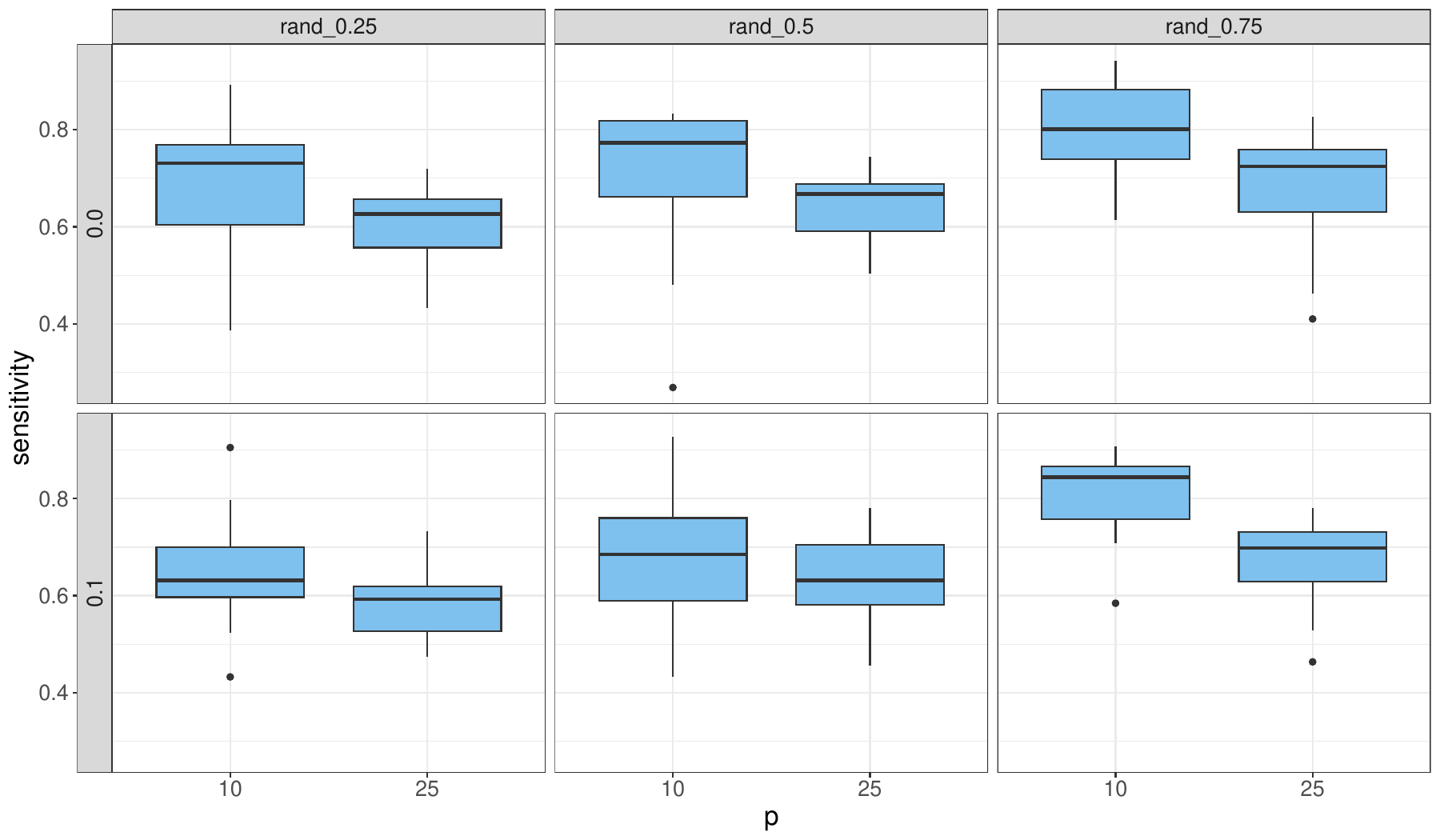}
        \caption{}
        \label{fig:sim_compsens_pois}
    \end{subfigure}
    \caption{Sensitivity of the zero recovery of the S-Bartlett by varying the missingness (cols) and the sparsity proportion (cols) for banded (a) and random (b) sparsity patterns.}
    
\end{figure}

\begin{figure}[H]
    \centering
  
    \begin{subfigure}[b]{.48\textwidth}
        \centering
       \includegraphics[width=0.99\linewidth]{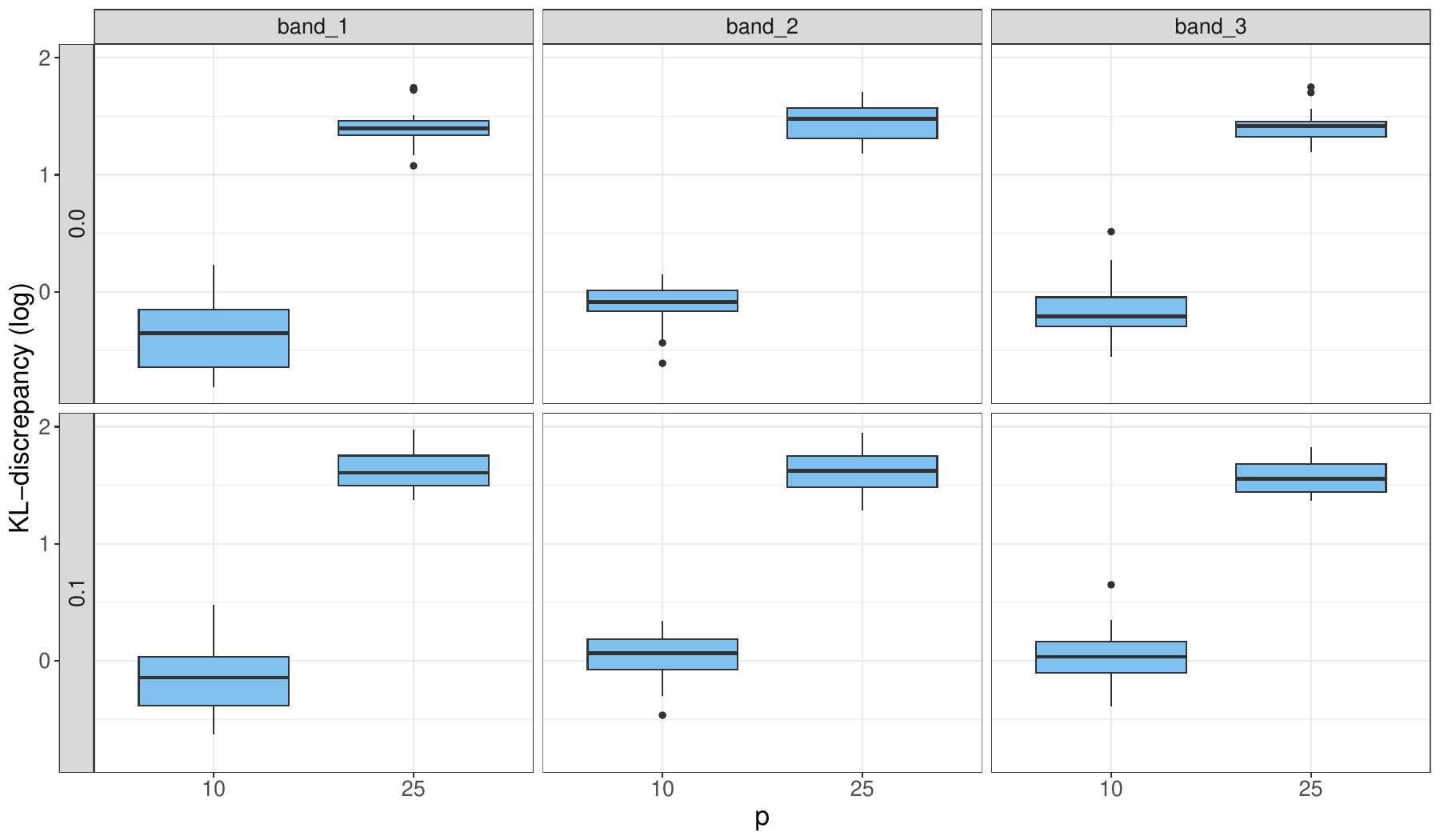}
        \caption{}
        \label{fig:sim_compKL_banded_pois}
    \end{subfigure}
    \begin{subfigure}[b]{.48\textwidth}
        \centering
         \includegraphics[width=0.99\linewidth]{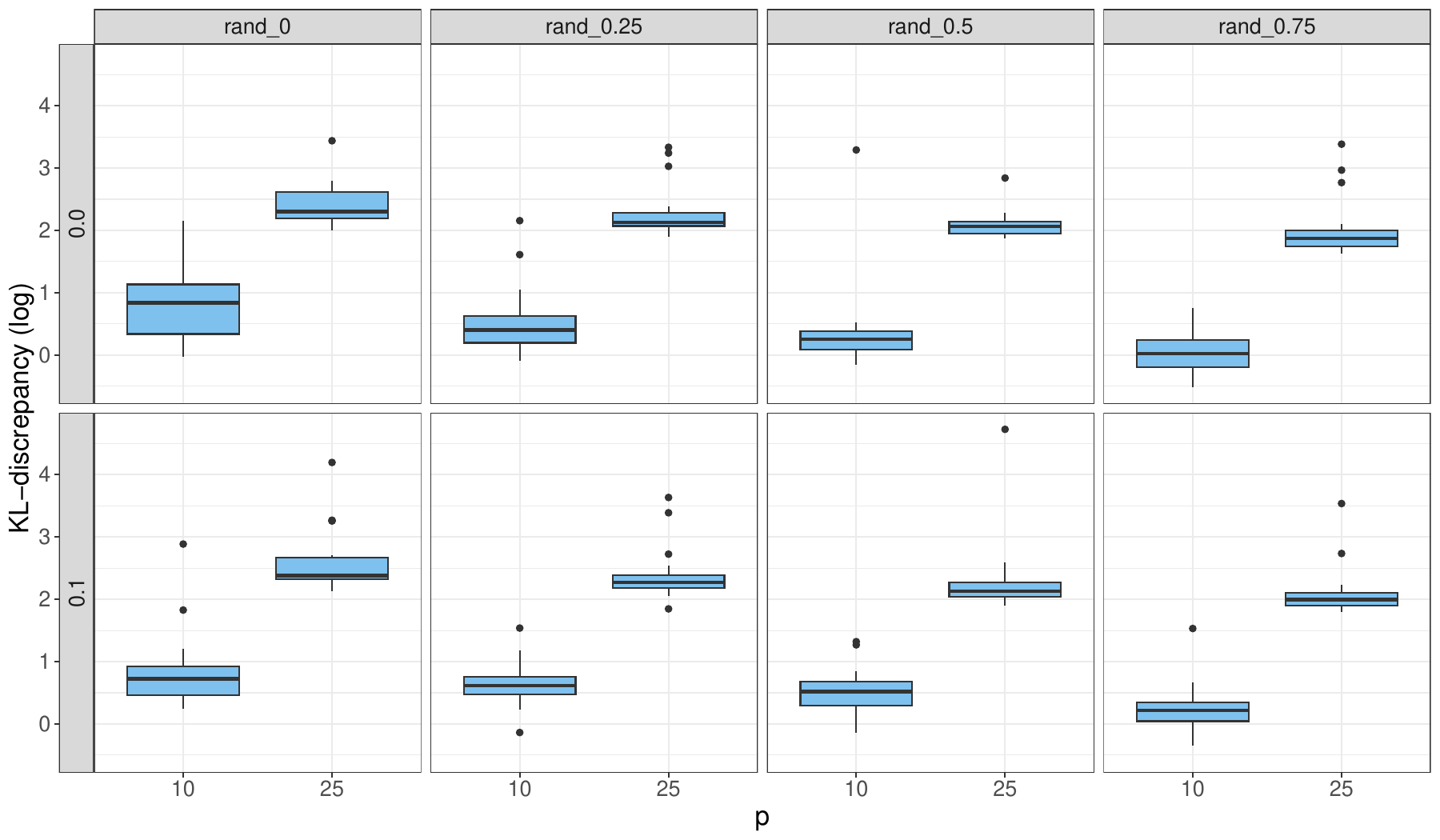}
        \caption{}
        \label{fig:sim_compKL_pois}
    \end{subfigure}
    \caption{KL-discrepancy between the true $\bLambda$ and the estimated $\hat{\bLambda}$ by S-Bartlett by varying the missingness (cols) and the sparsity proportion (cols) for banded (a) and random (b) sparsity patterns.}
\end{figure}

Figure \ref{fig:sim_compsens_banded_pois} and Figure \ref{fig:sim_compsens_pois} show the sensitivity distribution (across the replicas) for increasing outcome dimension obtained by the S-Bartlett by varying the number of missing data and the sparsity proportion, banded and random, respectively. The results highlight that the sensitivities are comparable to those obtained for the Normal likelihood -- provided the same scenario -- with slightly worsening performance as $k$ increases. Similarly, Figure \ref{fig:sim_compKL_banded_pois} and Figure \ref{fig:sim_compKL_pois} show the distribution of the KL-discrepancy, which yield to the same conclusions.

\newpage
\section{Further results for gene expression data}
\begin{figure}[H]
    \centering
    \begin{subfigure}[b]{.4\textwidth}
        \centering
        \includegraphics[width=0.95\linewidth]{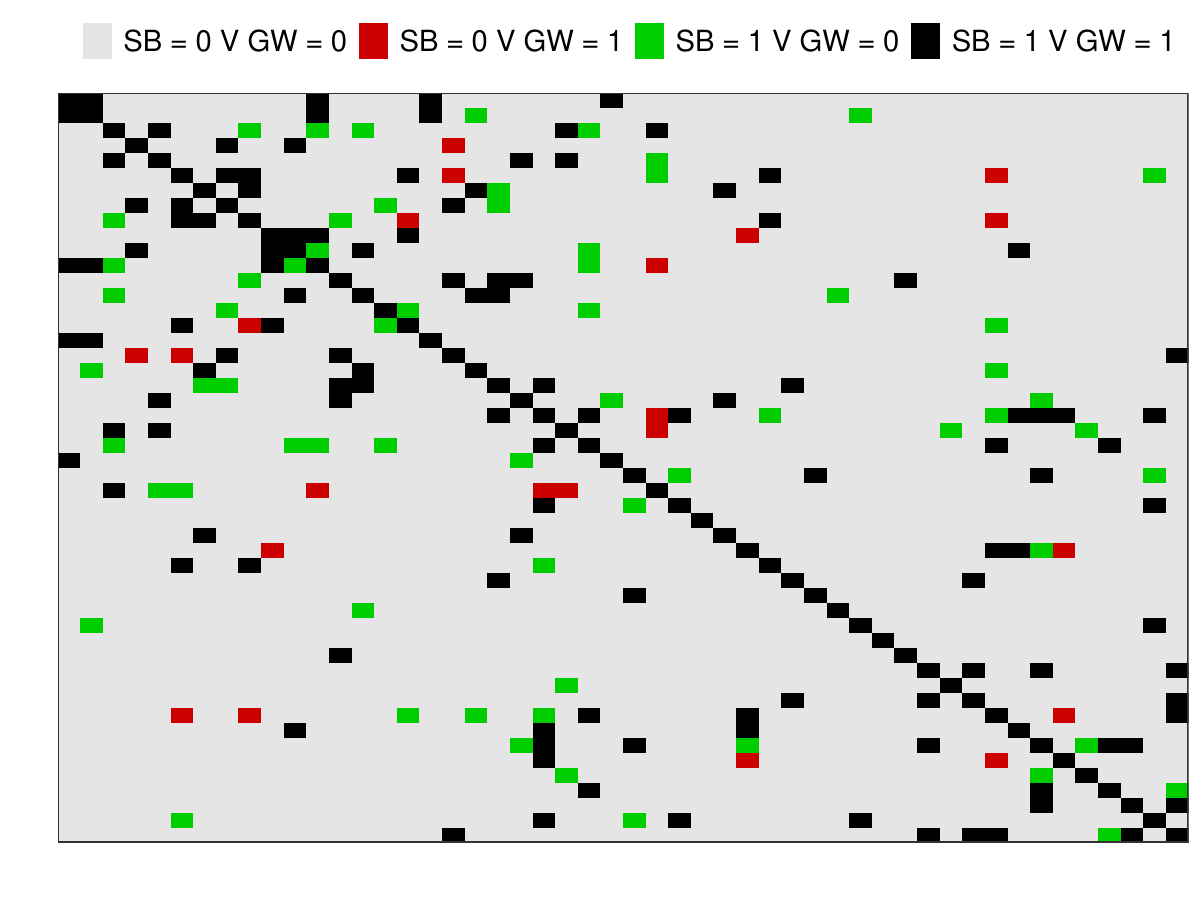}
        \caption{}
        \label{fig:compgene50}
    \end{subfigure}
    \begin{subfigure}[b]{.4\textwidth}
        \centering
       \includegraphics[width=0.95\linewidth]{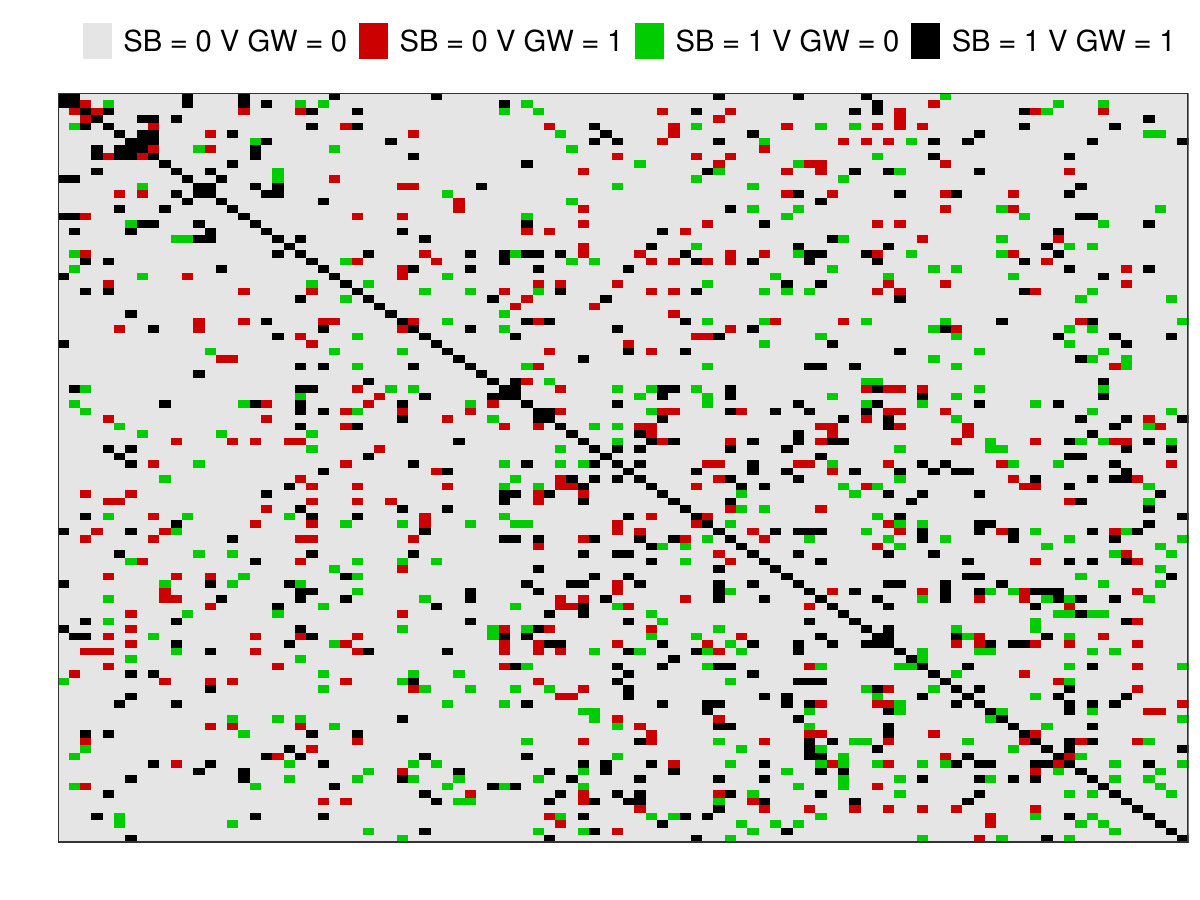}
        \caption{}
        \label{fig:compgene100}
    \end{subfigure}
    \caption{Comparison of the estimated $\bZ$ by the G-Wishart and the S-Bartlett prior for (a) $p=50$ and (b) $p=100$.}
\end{figure}

Figure \ref{fig:compgene50} and Figure \ref{fig:compgene100} show the differences between the estimated $\bZ$ when using the G-Wishart and the S-Bartlett prior for $p=50$ and $p=100$, respectively. On the one hand, the grey and black colors highlight agreement between the two methods, i.e. equivalent estimates of $\hat{z}_{jk} = 0$ and $\hat{z}_{jk} = 1$. On the other hand, there is disagreement between the two methods: the green color highlights cells that are turned on by the S-Bartlett but turned off by the G-Wishart, while the red color is for the opposite case. The percentage of mismatch is equal to 3.5\% of cells for $p=50$ and to 8\% for $p=100$.

\begin{figure}[H]
    \centering
    \begin{subfigure}[b]{.4\textwidth}
        \centering
        \includegraphics[width=0.95\linewidth]{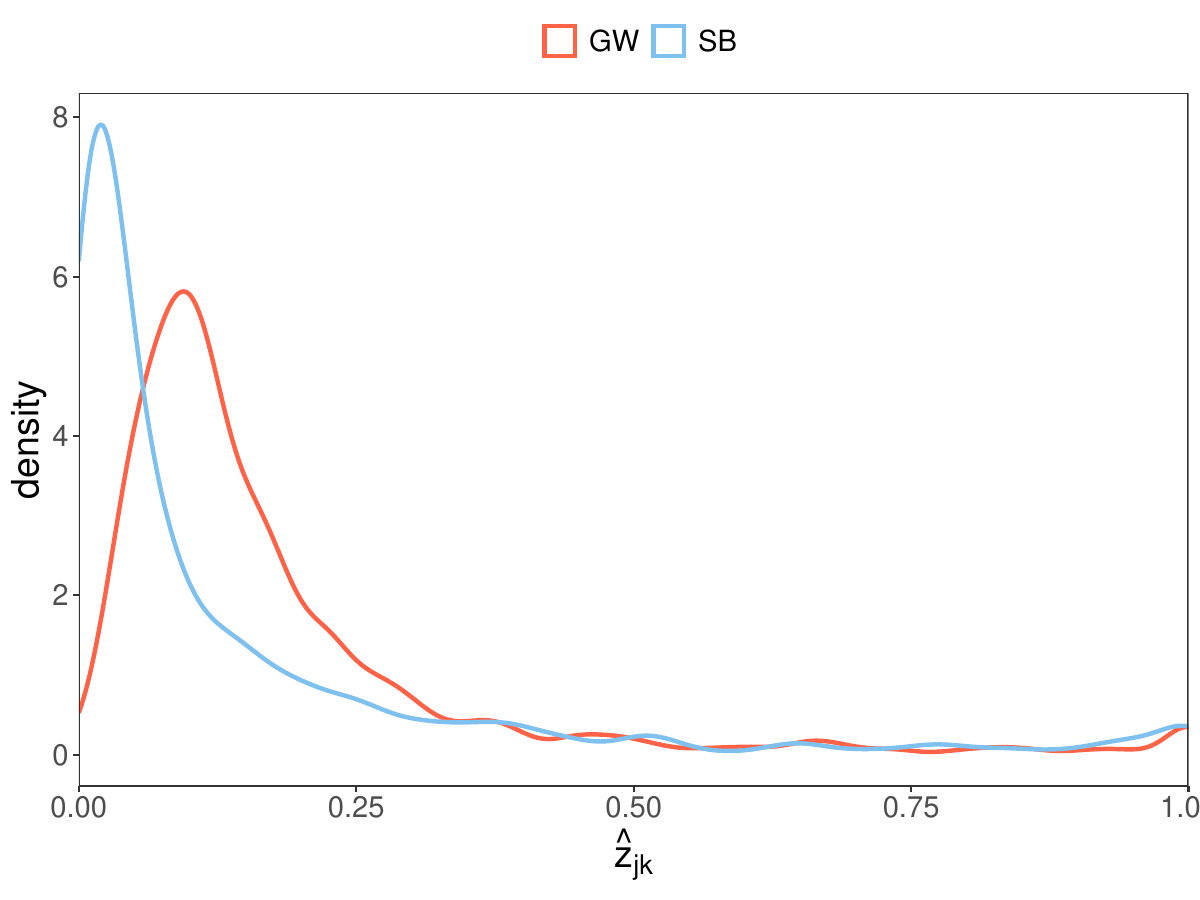}
        \caption{}
        \label{fig:compgenedistr50}
    \end{subfigure}
    \begin{subfigure}[b]{.4\textwidth}
        \centering
       \includegraphics[width=0.95\linewidth]{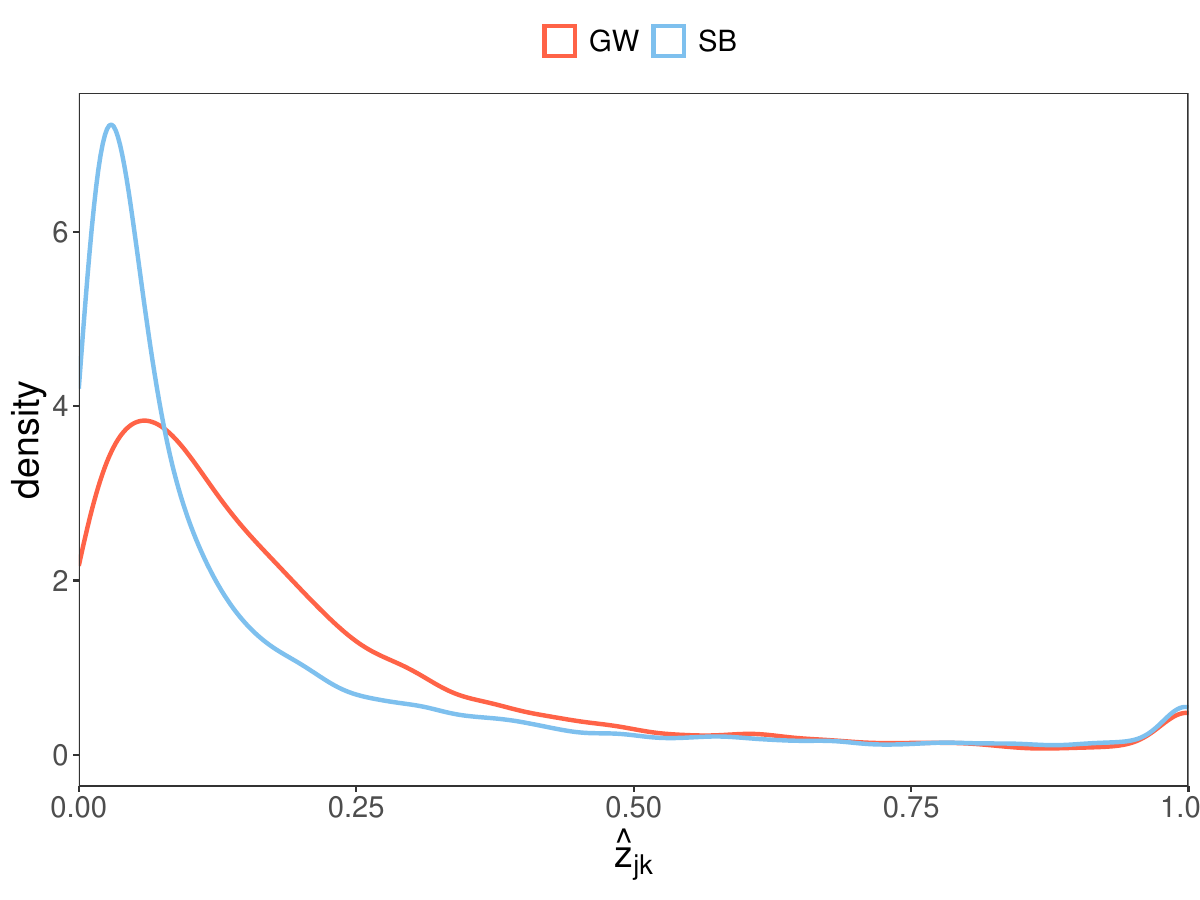}
        \caption{}
        \label{fig:compgenedistr100}
    \end{subfigure}
    \caption{Distribution of the estimated $\hat{z}_{jk}$ by the G-Wishart and the S-Bartlett prior for (a) $p=50$ and (b) $p=100$.}
\end{figure}

\end{document}